\documentclass{aa}   

\usepackage[comma,authoryear]{natbib}
\usepackage{graphicx}
\usepackage[varg]{txfonts} 
\usepackage{longtable}   

\newcommand{\eg}{{e.g.}}
\newcommand{\ie}{{i.e.}}
\newcommand{\ms}{m.s$^{\rm -1}$}
\newcommand{\kms}{km.s$^{\rm -1}$}
\newcommand{\msy}{m.s$^{\rm -1}$.y$^{\rm -1}$}
\newcommand{\msysq}{m.s$^{\rm -1}$.y$^{\rm -2}$}
\newcommand{\Mjup}{M$_{\rm Jup}$}

\newcommand{\vsini}{$v\sin{i}$}
\newcommand{\Msun}{M$_{\sun}$}
\newcommand{\Rsun}{R$_{\sun}$}
\newcommand{\Mstar}{M$_{\rm \star}$}
\newcommand{\bv}{$B-V$}
\newcommand{\msini}{$m_{\rm p}\sin{i}$}
\newcommand{\rhk}{log$R'_{\rm HK}$}
\newcommand{\safir}{S{\small AFIR}}
\newcommand{\sophie}{S{\small OPHIE}}
\newcommand{\harps}{H{\small ARPS}}

\newcommand{\elodie}{E{\small LODIE}}
\newcommand{\vega}{V{\small EGA}}

\begin{document}

   \title{Extrasolar planets and brown dwarfs around AF-type stars.
    \thanks{Based in part on observations made at Observatoire de Haute Provence (CNRS), France.}}

 \subtitle{X. {\bf The SOPHIE northern sample. Combining the SOPHIE and HARPS surveys to compute the close giant planet mass-period distribution around AF-type stars.}}

   \author{
     S. Borgniet \inst{1}
     \and
     A.-M. Lagrange \inst{1}
      \and
     N. Meunier \inst{1}
      \and
     F. Galland \inst{1}
      \and
     L. Arnold \inst{2}
      \and
     N. Astudillo-Defru \inst{3}
      \and
     J.-L. Beuzit \inst{1}
      \and
     I. Boisse \inst{4}
      \and
     X.~Bonfils \inst{1}
      \and
     F. Bouchy \inst{3,4}
      \and
     K. Debondt \inst{1}
      \and
     M. Deleuil \inst{4}
      \and
     X. Delfosse \inst{1}
      \and
     M. Desort \inst{1}
      \and
     R. F. D\'{i}az \inst{3}
      \and
     A. Eggenberger \inst{3}
      \and
     D.~Ehrenreich \inst{3}
      \and
     T. Forveille \inst{1}
      \and
     G. H\'{e}brard \inst{2,5}
      \and
     B. Loeillet \inst{5}
      \and
     C. Lovis \inst{3}
      \and
     G. Montagnier \inst{2,5}
      \and
     C. Moutou \inst{4,6}
      \and
     F. Pepe \inst{3}
      \and
     C.~Perrier \inst{1}
      \and
     F. Pont \inst{3}
      \and
     D. Queloz \inst{3,7}
      \and
     A. Santerne \inst{4}
      \and
     N. C. Santos \inst{8,9}
      \and
     D. S\'{e}gransan \inst{3}
      \and
     R. da Silva \inst{10,11}
      \and
     J. P. Sivan \inst{2}
      \and
     S. Udry \inst{3}
      \and
     A. Vidal-Madjar \inst{5}
             }
   \institute{
Univ. Grenoble Alpes, CNRS, IPAG, F-38000 Grenoble, France
\and
Observatoire de Haute-Provence, CNRS, Aix-Marseille Universit\'{e}, Institut Pyth\'{e}as UMS 3470, 04870 St Michel l'Observatoire, France
\and
Observatoire Astronomique de l'Universit\'{e} de Gen\`{e}ve, 51 Chemin des Maillettes, 1290 Versoix, Switzerland
\and
Aix Marseille Universit\'{e}, CNRS, LAM (Laboratoire d'Astrophysique de Marseille) UMR 7326, 13388 Marseille, France
\and
Institut d'Astrophysique de Paris, UMR 7095 CNRS, Universit\'{e} Pierre \& Marie Curie, 98bis boulevard Arago, 75014 Paris, France
\and
Canada-France-Hawaii Telescope Corporation, 65-1238 Mamalahoa Hwy, Kamuela, HI 96743, USA
\and
Cavendish Laboratory, J J Thomson Avenue, Cambridge CB3 0HE, UK
\and
Instituto de Astrof\'{i}sica e Ci\^{e}ncias do Espaço, Universidade do Porto, CAUP, Rua das Estrelas, 4150-762 Porto, Portugal
\and
Departamento de F\'{i}sica e Astronomia, Faculdade de Ci\^{e}ncias, Universidade do Porto, Rua do Campo Alegre, 4169-007 Porto,
Portugal
\and
ASI - Space Science Data Center (SSDC), Via del Politecnico snc, 00133 Rome, Italy
\and
INAF - Osservatorio Astronomico di Roma, via Frascati 33, 00078 Monte Porzio Catone, Rome, Italy
}
\offprints{simon.borgniet@obspm.fr}
   \date{Received date / Accepted date}

   
   \abstract
   {The impact of the stellar mass on the giant planet properties is still to be fully understood. Main-Sequence (MS) stars more massive than the Sun remain relatively unexplored in radial velocity (RV) surveys, due to their characteristics that hinder classical RV measurements.}
   {Our aim is to characterize the close (up to $\sim$2 au) giant planet (GP) and brown dwarf (BD) population around AF MS stars and compare this population to stars with different masses.}
   {We used the \sophie~spectrograph located on the 1.93m telescope at Observatoire de Haute-Provence to observe 125 northern, MS AF dwarfs. We used our dedicated \safir~software to compute the RV and other spectroscopic observables. We characterized the detected substellar companions and computed the GP and BD occurrence rates combining the present \sophie~survey and a similar \harps~survey.}
   {We present new data on two known planetary systems around the F5-6V dwarfs HD 16232 and HD 113337. For the latter, we report an additional RV variation that might be induced by a second GP on a wider orbit. We also report the detection of fifteen binaries or massive substellar companions with high-amplitude RV variations or long-term RV trends. Based on 225 targets observed with \sophie~and/or \harps, we constraint the BD frequency within 2-3 au around AF stars to be below 4\%~(1$\sigma$). For Jupiter-mass GP within 2-3 au (periods $\leq 10^{3}$ days), we found the occurrence rate to be $3.7_{-1}^{+3}$\%~around AF stars with masses $< 1.5$ \Msun, and to be $\leq 6$\%~(1$\sigma$) around AF stars with masses $> 1.5$ \Msun. For periods smaller than 10 days, we find the GP occurrence rate to be below 3 or 4.5\%~(1$\sigma$), respectively. Our results are compatible with the GP frequency reported around FGK dwarfs and are compatible with a possible increase of GP orbital periods with the stellar mass as predicted by formation models.}
   {} 
   \keywords{Techniques: radial velocities - Stars: early-type - Stars: planetary systems - Stars: variable: general}

   \maketitle

\section{Introduction}\label{sect:intro}

More than three thousands exoplanets and brown dwarfs (BD) have now been confirmed\footnote{\texttt{http://exoplanet.eu}, \cite{schneider11}.}, while thousands of other {\it Kepler} candidates await confirmation. Most of these substellar companions are close ($\leq$5-10 au) from their host star and have been detected via radial velocity (RV) and transit surveys \citep[see \eg][]{borucki10,howard11,fressin13,mayor14}. Giant, gaseous planets are at the core of the planetary systems as they carry most of their mass. These close giant planets (GP) have revealed an unexpected diversity in terms of both orbital (period, eccentricity, inclination) and physical (mass, composition) properties. This variety underlines the complexity of the different processes and interactions that shape the GP population, both in terms of formation, migration, dynamical interactions or influence of the primary host star.\\

While our understanding of the GP formation and evolution mechanisms has dramatically improved over the past twenty years, the influence of the stellar properties on the GP distribution remains a key topic. For example, it is now well-established that the giant planet (GP) frequency increases with the stellar metallicity for solar-like, Main-Sequence (MS) FGK dwarfs \citep{santos04,fischer05}. 

However, the impact of the stellar host mass on its companion properties is still not fully understood. The core-accretion formation process, that is believed to be at the origin of most of the RV and transit planets, predicts an increase of the GP frequency with the stellar mass \citep[up to 3-3.5 \Msun][]{kennedy08}. This was confirmed by comparing the GP occurrence rates derived from RV surveys of low-mass M dwarfs \citep[see \eg][]{bonfils13} and of solar-mass FGK dwarfs \citep[see \eg][]{cumming08}. Subgiant and giant stars seem to confirm this trend for higher stellar masses, as higher GP occurrence rates than around solar-mass dwarfs have been reported for these evolved stars \citep[see \eg][]{johnson10,reffert14}. However, the actual mass and origin of these evolved stars is subject to an ongoing controversy, as several studies argue that these giants do not significantly differ from solar-type stars in terms of masses, and are consequently their descendants \citep[instead of being the descendants of AF MS stars; see the list of references in][]{borgniet17}.

AF stars with well-established masses above 1 \Msun~on the Main Sequence are generally rejected from RV surveys due to their specific characteristics (fewer spectral lines and faster rotation) that hinder classical RV measurements. A recent survey has targeted chemically peculiar Ap stars, which exhibit lower rotational velocities and more narrow spectral lines than typical AF MS dwarfs \citep{hartmann15}.\\

Our group developed twelve years ago the \safir~tool \citep[Software for the Analysis of the Fourier Interspectrum Radial velocities][]{galland05a} specifically dedicated to derive RV measurements of AF MS stars as precise as possible. For one target, \safir~cross-correlates each observed spectrum with a reference spectrum built from all the acquired spectra instead of a binary mask, allowing to reach a better RV precision \citep[closer to the photon noise limit defined by][]{bouchy01}. Even if this RV precision remains intrinsically degraded compared to cooler and/or slower rotating stars, we developed \safir~(as well as other specific tools) to demonstrate that close ($\lesssim$3 au), massive ($\geq$ 1 \Mjup) GP and BD companions can indeed be searched for around AF MS stars \citep{galland05b,galland06,lagrange09,meunier12}.

From 2005 to 2016, we carried out a survey targeting GP and BD orbiting within $\sim$3 au around southern AF MS dwarfs with the \harps~spectrograph \citep[High-Accuracy Radial velocity Planet Searcher, see][]{pepe02} on the 3.6 m ESO telescope at La Silla Observatory in Chile. During this survey, we {\it i}) detected three new GP around two F5-6V stars \citep{desort08}, {\it ii}) showed that GP more massive than Jupiter and BD are often detectable at periods up to a thousand days around MS AF dwarfs more massive than the Sun \citep{lagrange09}, and {\it iii}) derived the first constraints on the occurrence rates of such companions \citep[][hereafter Paper IX]{borgniet17}. Meanwhile, we carried out a survey of northern AF MS stars with the \sophie~echelle fiber-fed spectrograph \citep[][]{bouchy06}, which is the successor of the \elodie~spectrograph \citep{baranne96} on the 1.93~m telescope at Observatoire de Haute-Provence in France. Most of the observations were made in the course of the SP4 program within the \sophie~consortium \citep{bouchy09}. The first results of our \sophie~survey, regarding the complex RV variations of HD~185395 and the detection of a $\sim$3~\Mjup~GP orbiting with a $\sim$320-day period around HD~113337, were presented in \cite{desort09} and \cite{borgniet14}, respectively.\\

\begin{figure*}[ht!]
  \centering
\includegraphics[width=0.9\hsize]{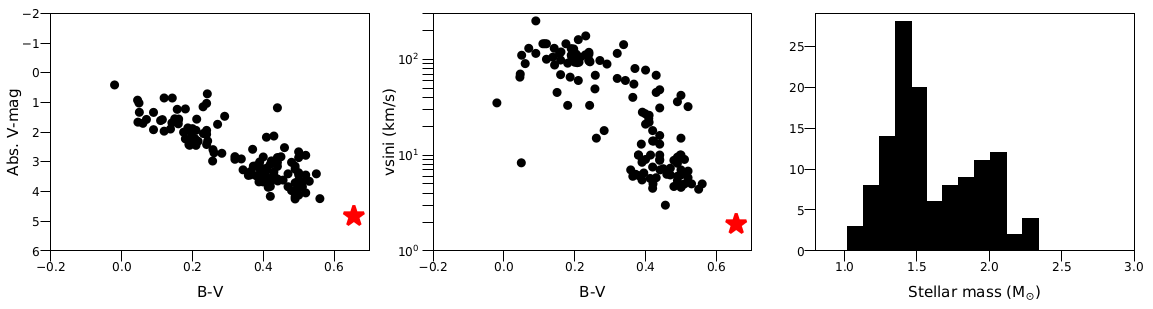}
\caption{Main physical properties of our sample. {\it Left}: HR diagram of our sample, in absolute $V$-magnitude vs \bv. Each black dot corresponds to one target. The Sun is displayed ({\it red star}) for comparison. {\it Middle}: \vsini~vs \bv~distribution. {\it Right}: Mass histogram of our sample (in \Msun).} 
\label{sample}
\end{figure*}

This paper first presents our \sophie~survey and its results, and then presents a statistical analysis combining our \sophie~and \harps~targets from Paper IX to make a single, global study. The \sophie-\harps~combination is possible here because: {\it i}) the \harps~and \sophie~instruments work on the same principles; {\it ii}) they have a similar instrumental RV precision (less than 1 \ms~for \harps~versus a few \ms~for \sophie); {\it iii}) our surveys have the same overall characteristics; and {\it iv}) we use exactly the same reduction and analysis that we made for our \harps~sample in Paper IX. In Sect.~\ref{sect:survey}, we describe our \sophie~sample and observations and give a brief outline of our computational tool, our observables and our RV variability diagnosis. Sect.~\ref{sect:det} is dedicated to the detection and characterization of giant planets and other long-term companions within our \sophie~survey. In Sect.~\ref{sect:analysis} we make a global analysis of the combined \sophie~+ \harps~sample: {\it i}) we characterize the RV intrinsic variability of our targets, {\it ii}) we derive the search completeness of our combined survey, and {\it iii}) we combine the companion detections and the detection limits to compute the close companion (GP and BD) occurrence rates and estimate their (mass,~period) distribution around AF MS stars. We conclude in Sect.~\ref{sect:conclu}.

\section{Description of the \sophie~survey}\label{sect:survey}

\begin{figure*}[ht!]
  \centering
\includegraphics[width=0.9\hsize]{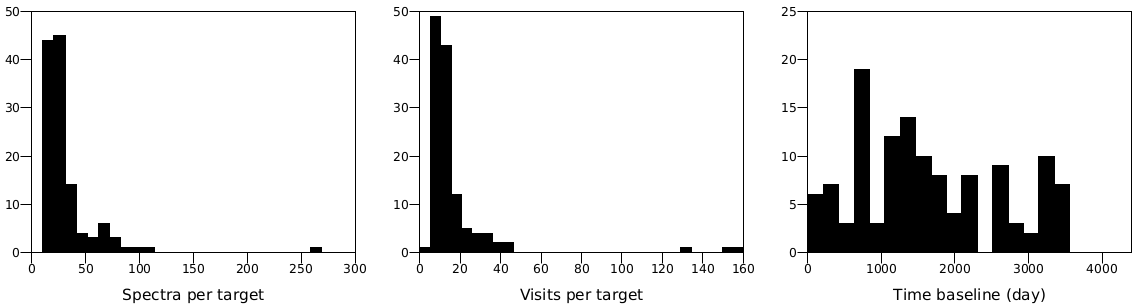}
\caption{Observation summary. {\it Left}: Histogram of the spectrum number per target. {\it Middle}: Histogram of the separate observation epoch number per target. {\it Right}: Histogram of the time baselines.} 
       \label{sample_obs}
\end{figure*}

\subsection{Sample}\label{subsect:sample}

Our \sophie~sample is made of 125 AF MS dwarf stars with spectral types in the range A0V to F9V. We detail the physical properties of our sample in Appendix~\ref{app:sample}. We selected our targets using the same selection process as for our \harps~survey \citep[see more details in][Paper IX and Appendix~\ref{app:select}]{lagrange09}. Briefly, our selection relies on the following criteria: {\it i}) the spectral type; {\it ii}) the distance to the Sun, and {\it iii}) the exclusion of known SB2 binary, chemically peculiar stars, or confirmed $\delta$~Scuti or $\gamma$~Doradus pulsators. We ended up with 125 targets with \bv~in the range -0.02 to 0.56, \vsini~in the range 3 to 250 \kms, and masses in the range 1 to 2.5 \Msun~(Fig.~\ref{sample}). Note that the exclusion of known pulsators results in a visible dichotomy between our $\sim$A- and $\sim$F-type targets.

\subsection{Observations}\label{subsect:obs}

We observed our 125 \sophie~targets mainly between November 2006 and April 2014. We made some additional observations later (2015-2016) to complete our survey for the targets with the less data and to further monitor the most interesting targets. We acquired the \sophie~spectra in high-resolution mode ($R \sim$$75000$), in the 3872-6943 \AA~wavelength range. As done with our \harps~targets, we adapted the exposure times depending on the observing conditions and on the target magnitude in order to get a high signal-to-noise ratio (S/N), of at least 100 at $\lambda = 550$ nm. We made most of the exposures in the simultaneous-Thorium mode, for which the \sophie~A fiber is centered on the target while the B fiber is fed by a Thorium lamp. This allows to follow and correct for a potential instrumental RV drift induced by local temperature or pressure variations, if needed. For very bright targets (very small exposure times), we acquired the spectra in the \sophie~standard spectroscopy mode (which is appropriate for very bright targets), for which the A fiber is centered on the target while the B fiber is not illuminated. Our observing strategy and analysis are fully detailed in Paper IX. Briefly, we acquire at least two consecutive spectra at each epoch/visit during several consecutive nights to estimate the short-term stellar jitter of our $\sim$F-type targets, whereas we acquire multiple consecutive spectra per epoch for our $\sim$A-type targets, to mitigate the effect of pulsations.

The median time baseline of our \sophie~survey is 1448 days or $\sim$4 years (the mean time baseline being 1640 days or 4.5 years), with a median spectrum number per target of 23 (36 on average) acquired during a median number of 11 visits (17 on average, Fig.~\ref{sample_obs}). 

\subsection{Observables}\label{subsect:data}

As done for our \harps~sample, we compute the RV and, whenever possible, the line profiles and the chromospheric emission in the Calcium lines of our targets with our software \safir~\citep{chelli00,galland05a}. The main principles of \safir~are fully detailed in Paper IX. Our main observables are the RV, the spectrum cross-correlation function (CCF) and its full width at half maximum (FWHM), the bissector velocity span \citep[hereafter BIS,][]{queloz01} and the \rhk. Our main (but not the only) diagnosis to classify the RV variability of our targets and distinguish between its possible sources is the (RV,~BIS) diagram. See Paper IX for a fully detailed description of our various observables and of our variability criteria and diagnosis. Very briefly, we classify our targets as RV variable if the total RV amplitude and the RV standard deviation are at least six (respectively two) times larger than the mean RV uncertainty. A RV-BIS anti-correlation is most likely the indication of stellar magnetic activity, a RV-BIS vertical diagram indicates pulsations and a flat RV-BIS diagram points toward companionship; CCF distortions and/or RV-FWHM correlation might indicate the presence of a SB2 or SB1 binary. More detailed results on our target observables can be found in Appendix~\ref{app:sample}.

Note that we quadratically add 5 \ms~of systematic error to the \sophie~uncertainties to take into account the \sophie~instrumental error \citep{diaz12,bouchy13,borgniet14}. Such an error may be slightly excessive for the latest \sophie~measurements (see below), but we prefer to stay conservative and keep the same uncertainty for all our data.

\subsection{\sophie+}

In June 2011 (BJD 2455730), \sophie~was updated (nicknamed \sophie+) by implementing octagonal fibers at the fiber link \citep{bouchy13}. Prior to this update, the \sophie~RV were known to exhibit a systematic bias induced by the insufficient scrambling of the old fiber link. This bias was called the ``seeing effect'' as seeing variations at the fiber entrance were causing RV variations \citep{boisse11b}. One way to correct for this bias is to measure the RV in the blue and red halves of each \sophie~spectral order and compute their difference $\delta_{\rm RV}$, which should be correlated to the original RV if affected by the seeing effect, as detailed by \cite{diaz12}. We applied the same correction in \safir, computing $\delta_{\rm RV}$ for all spectra acquired prior to June 2011, and correcting the RV from the $\delta_{\rm RV}$ correlation if needed (\ie~if the Pearson's correlation coefficient was above 0.5 in absolute value). This happened for sixteen of our targets for which the $\delta_{\rm RV}$ correlation was $>$ 0.5.

\section{Detected companions in the \sophie~survey}\label{sect:det}

\subsection{Giant planets}\label{subsect:GP}

In this section, we present two planetary systems with three GP, either confirmed or probable candidates. The host stars’ fundamental properties are detailed in Table~\ref{table:stell_param}, and the orbital parameters of the detected and/or candidate GP in Table~\ref{table:orbit_param}. The RV and line profile data of these systems are showed in details in Appendix~\ref{app:details}.

\subsubsection{The HD~113337 system}\label{subsubsect:hd113}

\paragraph{Summary of first planet detection --}

We reported the detection of a giant planet orbiting around HD~113337 and characterized it in \cite{borgniet14}, hereafter BO14. HD~113337 (HIP~63584) is a F6V star with a 1.41 \Msun~stellar mass \citep{allende99}, a slight infrared excess attributed to a cold debris disk \citep{rhee07}, and may be relatively young (possible age of $150 \pm 100$ Myr, see BO14). Based on 266 \sophie~RV measurements spanning six years, we reported clear $\sim$320-day periodic RV variations along with a longer-term, quadratic-shaped RV drift and a flat (RV,~BIS) diagram. We showed that the periodic RV variations are induced by a \msini~$\sim$2.8 \Mjup~giant planet orbiting around HD~113337 with a 324-day period on an eccentric orbit ($e \simeq 0.46$). Based on weak long-term variations observed in the BIS and FWHM, we finally supposed that the long-term quadratic RV variation was more probably induced by a stellar activity cycle than by an additional companion.

\paragraph{New \sophie~data --}

As a target of great interest, we continued to observe HD~113337 with \sophie~in 2014-2016, adding 35 additional spectra to our database and consequently expanding the time baseline by $\sim$3 years, up to 9.2 years (3369 days). We display the main \sophie~spectroscopic observables for this target in Fig.~\ref{hd113337}. Most interestingly, the new RV data exhibits a clear rebound on the long-term scale around Julian Day (hereafter JD) $\sim$2457000. It thus means that the long-term RV variations cannot longer be modeled by a quadratic fit anymore and could possibly be periodic. We computed the RV periodograms according to the Lomb-Scargle definition \citep{scargle82,press89}, and also with the CLEAN algorithm \citep[which deconvolves the Lomb-Scargle periodogram from the window function, see][]{roberts87}. Both periodograms exhibit two clear peaks (Fig.~\ref{hd113337}, top row): {\it i}) one at $\sim$320 days, corresponding to the already known GP (hereafter HD~113337b); and {\it ii}) another one at $\sim$3200 days, corresponding to the long-term variations.\\

The BIS data has a 24.5 \ms~dispersion and does not show any high-amplitude variations. However, a slight long-term variation may be present with a BIS extremum between BJD~2455600 and BJD~2456500, approximately (Fig.~\ref{hd113337}, second row). Except for two peaks at $\sim$2 and $\sim$3 days that are probably induced by the stellar rotation, the BIS Lomb-Scargle periodogram does not exhibit any significant signal for periods up to 1000 days. However it shows power between 2000 and 4000 days, \ie~in roughly the same period range as the observed RV long-term variations. When looking at the BIS CLEAN periodogram, the long-term peak is centered at a $\sim$2000-day periodicity (\ie~quite distinct from the $\sim$3200-day RV periodicity) and is no longer the highest peak in the periodogram (the highest peak being at $\sim$2.5 days). The (RV,~BIS) diagram is nearly flat (with a $−0.12 \pm 0.02$ linear slope and a Pearson correlation coefficient of -0.32), thus ruling out stellar activity as the only source of the RV variations.\\

The FWHM time series exhibits a high-amplitude ($\sim$250 \ms) long-term variation, with a minimum between BJD~2455700 and BJD~2456200 (Fig.~\ref{hd113337}, third row). In the same way as the BIS, the FWHM Lomb-Scargle periodogram exhibits power between 2000 and 4000 days, and the FWHM CLEAN periodogram exhibits a clear, single peak around 2200 days. There is no other significant signal at shorter periods; the peaks present in the Lomb-Scargle periodogram around 200-400 days and at $\sim$30 days are aliases induced by the temporal sampling (see window function). The Pearson coefficients for the (RV,~FWHM) and (BIS,~FWHM) couples are 0.25 (no correlation) and -0.49 (anti-correlation), respectively. This suggests a common origin for the long-term (high-amplitude) FWHM and (low-amplitude) BIS variations (possibly an activity cycle) while giving further credence to another origin to the RV variations.\\

HD~113337 does not show any chromospheric emission in the Calcium H and K lines \citep{borgniet14}. The \rhk~has a mean value of -4.76 and does not exhibit any significant signal.

\paragraph{A second GP around HD~113337 ?}

Given the new \sophie~RV data, one can try to fit HD~113337 RV variations with a 2-planet Keplerian orbital model (Fig.~\ref{hd113337}). To perform the Keplerian fits, we used the {\it yorbit} software \citep{segransan11}, as we did in our previous studies (see \eg~BO14 and Paper IX). Our two-planet best solution gives orbital parameters for planet b that are in agreement with the values that we derived in BO14 in terms of period ($323 \pm 1$ days against $324 \pm 2$ days, respectively) and minimal mass ($3 \pm 0.3$ \Mjup~against $2.8 \pm 0.3$ \Mjup, respectively). Interestingly, the eccentricity of planet b is significantly reduced by going from a 1-planet to a 2-planet solution (from $0.46 \pm 0.04$ down to $0.36 \pm 0.03$, respectively). If considering that the RV long-term variations are induced by a second GP (hereafter planet c), our best model would correspond to a period of $\sim 3265 \pm 134$ days and a $6.9 \pm 0.6$ \Mjup~minimal mass on a slightly eccentric (e $\simeq$ 0.2) orbit. The RV residuals of the 2-planet Keplerian fit show a remaining dispersion of $\sim$23 \ms~(Fig.~\ref{hd113337}, fifth row), with a significant anti-correlation with the BIS (Pearson coefficient of -0.44). The periodograms of the RV residuals mainly show small peaks at $\sim$2 and $\sim$4 days, as well as at $\sim$90 days. None of these periodicities can be properly fitted with {\it yorbit}. We attribute the short-period peaks to the stellar rotation as they are also present in the BIS periodogram. The origin of the remaining $\sim$90-day peak is less clear, but a similar peak is present in the BIS periodograms (Fig.~\ref{hd113337}, second row). Thus, we do not find any significant sign of a third companion-induced periodicity. The orbital parameters we deduced from the 2-planet Keplerian fit are fully detailed in Table~\ref{table:orbit_param}.\\

While the long-term RV variation can be well modeled by a Keplerian fit corresponding to a second GP, other possible origins can also be considered, due to the presence of FWHM (and, on a much smaller scale, BIS) long-term variations. We review below different hypothesis and their pros and cons (second GP included), and explain why we consider the second GP hypothesis the most convincing one:
\begin{enumerate}
\item An activity cycle. Magnetic activity cycles are characterized by the rise and fall of the number of magnetically active structures on the stellar photosphere (such as dark spots and bright faculae) over durations much longer than the stellar rotation period. They can induce RV and/or line profile and/or \rhk~variations. Activity cycles with associated timescales of a few years are relatively common among FGK MS dwarfs \citep[see \eg][]{schroeder13}. For example, reconstructed solar RV exhibit a peak-to-peak RV amplitude of 8-10 \ms~over solar cycle 23 \citep[][]{meunier10}. Given that HD~113337 \vsini~of 6 \kms~is larger than \sophie~instrumental resolution ($\sim$4 \kms), magnetic stellar activity should induce both RV and line profile (BIS and/or FWHM) variations \citep[see \eg][and Paper IX]{desort07,lagrange09}.

In this hypothesis, an activity cycle with a duration of the order of the candidate planet c period would explain both the FWHM and BIS long-term variations (and their correlation) as well as the RV long-term variations. We consider though that this interpretation suffers an important weakness: the RV corrected from the Keplerian fit of planet b only and the FWHM are not correlated (Pearson's coefficient $<$ 0.1, see Figs.~\ref{hd113337} and~\ref{oc113337}). This is due to the RV and FWHM long-term variations being significantly shifted in phase (considering a period of $\sim$3265 days from the 2-planet Keplerian fit). A phase shift of 15 to 30\degr~between RV, BIS and FWHM activity-induced variations can be expected in the case of short-term stellar activity linked to the stellar rotational period \citep[see \eg][who show that an active region crossing the visible stellar disk induces a RV shift and a line profile deformation with a slight phase shift]{santos14,dumusque14}. However, such a phase shift cannot be present in the case of a long-term activity cycle, where the RV and line profile long-term variations are induced by the changing active region filling factor of the stellar disk along the cycle \citep{borgniet15}, being thus correlated \citep{lovis11}. In the case of HD~113337, we derived the Pearson’s coefficient of the RV residuals of planet b versus FWHM while shifting in phase the FWHM along the $\sim$3265-day period deduced from the 2-planet Keplerian fit (Fig.~\ref{oc113337}). The highest correlation ($\sim$0.65) is reached for a phase shift of $\sim$0.45 (or 1470 days) while the highest anti-correlation ($\sim -0.65$) is reached for a phase shift of $\sim$0.7 (or 2290 days). While we still do not understand everything about stellar magnetic activity cycles, we consider that the observed shift between the RV and FWHM long-term variations is a strong argument against the activity cycle origin for the RV long-term variations, given our present knowledge \citep{lovis11,dumusque11}. We note that long-period RV planets have already been distinguished from activity cycles by looking for the presence of a phase shift between the RV and the \rhk~taken as an activity proxy \citep{wright16}.

\begin{figure}[ht!]
  \centering
\includegraphics[width=1\hsize]{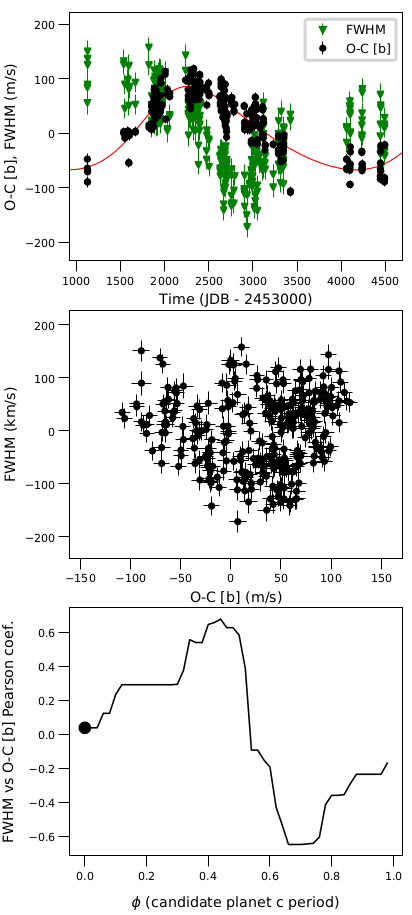}
\caption{{\it Top}: HD~113337 RV residuals from the Keplerian fit of planet b only vs time ({\it black dots}) and HD~113337 FWHM vs time ({\it green triangles}, same scale as RV). The Keplerian fit of candidate planet c is overplotted to the RV residuals of planet b ({\it red}). The average FWHM ($\sim$13.7 \kms) has been set to 0 for clarity. {\it Middle}: FWHM vs RV residuals of planet b. {\it Bottom}: Pearson’s correlation coefficient of RV residuals of planet b vs FWHM for different FWHM phase shifts, considering a $\sim$3265-day period ({\it black solid line}). The actual Pearson’s coefficient (for no FWHM phase shift) is displayed as a black dot.}
\label{oc113337}
\end{figure}

\item A faint, low-mass stellar companion to HD~113337 in a (nearly) pole-on configuration could be just bright enough to slightly blend the primary spectrum, and thus induce the long-term FWHM and (low-amplitude) RV variations. HD~113337 \vsini~is quite low for a F6V spectral type \citep[see][]{borgniet14}, meaning that the system could well be seen inclined (assuming that the star rotates in the same plane as the planetary system). Furthermore, HD~113337 is host to a cold debris disk that is most likely also see inclined (K. Su, {\it private comm.}). However, in such a configuration, the RV and FWHM companion-induced variations would also have to be correlated (see Sect.~\ref{subsubsect:hd191195}), which is not the case here. Moreover, given the $\sim$7 \Mjup~minimal mass of the second candidate companion, the system inclination would have to be smaller than 3\degr~to allow for an actual companion mass of $\sim$150 \Mjup, which makes it quite unprobable statistically.bStellar mass-luminosity relations \citep{baraffe98} taken for an age between 120 and 500 Myr show that a companion mass of 150 \Mjup~would translate into a contrast of 6 magnitudes with the primary in the $H$-band, and so even greater in the $V$-band. This would make the effect of the companion light on the primary spectrum completely negligible (especially when using a reference built from the primary spectra for the cross-correlation). Thus, we find the hypothesis of a low-mass stellar companion difficult to support given the observations.

\item Planet c. The RV and FWHM long-term variations are phase-shifted, which is not coherent with a magnetic activity cycle (see above). Furthermore, if plotting the BIS versus the RV residuals of planet b (Fig.~\ref{hd113337}, fourth row), we find a still nearly flat diagram with a linear slope of $−0.14 \pm 0.02$ and a correlation coefficient of -0.29. These values are very close to those we find for the (RV,~BIS) diagram (see above), meaning that the horizontal spread in the (RV,~BIS) diagram (characteristic of the presence of a companion) may well be induced by more than one planet. On the contrary, the RV residuals of the 2-planet fit plotted versus the BIS show a linear slope of $−0.49 \pm 0.05$ and an enhanced anti-correlation (-0.44, see above), meaning that the remaining RV dispersion is likely induced by stellar activity. We consider that the presence of a second GP in the system is thus a very convincing explanation for the RV long-term variations and the shapes of the RV and RV residuals versus BIS diagrams. The additional presence of a long-term activity cycle would then explain the long-term FWHM and BIS variations, and the anti-correlation present between the RV residuals and the BIS. 
\end{enumerate}
To conclude here, we decided to classify HD~113337 as a candidate 2-planet system (planet b being confirmed and planet c still a candidate).

\paragraph{Additional remarks --}
In addition to the \sophie~observations, we are carrying out a multi-technique (optical interferometry, disk imaging, and deep imaging of the outer environment) study of the HD~113337 system. The aim of this study is to bring as many constraints as possible on the system characteristics (both in terms of stellar fundamental parameters, age, inclination, etc) and to fully cover the companion (mass, separation) space. This will be the topic of a forthcoming, dedicated paper.

\subsubsection{The giant planet around HD~16232}\label{subsubsect:hd16232_GP}

HD~16232 (30 Ari B, HIP~12184) is a F6V dwarf star around which \cite{guenther09} reported the detection of a substellar companion of $\sim$10 \Mjup~minimal mass, based on RV observations. This companion was further characterized thanks to new RV measurements \citep[for a total of 110 spectra; see][hereafter K15]{kane15}. These authors made a new estimation of the companion orbital parameters, making it of planetary nature, and additionally detected a long-term linear RV trend likely induced by a distant stellar companion. The Keplerian+linear fit performed by K15 thus gives \msini~$= 6.6 \pm 0.9$ \Mjup, $P = 345.4 \pm 3.8$ days and $e = 0.18 \pm 0.11$ (corresponding to a $1.01 \pm 0.01$ au sma) for HD~16232b.\\

\begin{figure}[t!]
\centering
\includegraphics[width=1\hsize]{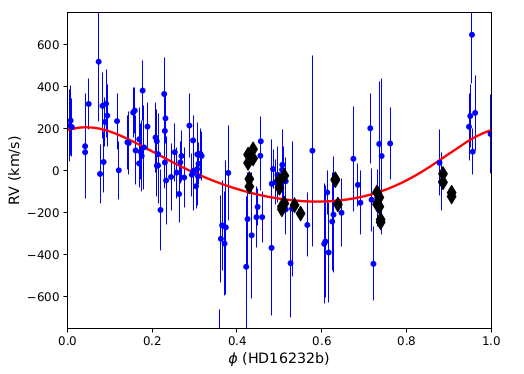}
\caption{HD~16232 phased RV. RV from K15 are displayed as blue dots while our \sophie~RV are displayed as black diamonds. The data are phased along the orbit of HD~16232b according to the Keplerian fit of K15 RV in Table~\ref{table:orbit_param} (the corresponding best model is displayed in red).}
\label{hd16232ph}
\end{figure}

As part of our \sophie~survey, we acquired 27 spectra on HD~16232, covering a 1893-day time baseline. These data are enough to allow us to clearly see the long-term RV trend firstly detected by K15 in our \sophie~RV (Fig.~\ref{hd16232}, third row). Furthermore, we find that the \sophie~BIS versus RV diagram looks composite, with an horizontal spread indicative of a companion and a vertical spread induced by stellar activity and/or low-level pulsations. The \sophie~BIS and FWHM do not show any significant signal. We characterize the binary companion at the origin of the long-term RV trend in more details in Sect.~\ref{subsubsect:hd16232_BIN}. However, we do not detect any other signal than the binary-induced trend in our \sophie~RV (Fig.~\ref{hd16232}, sixth row). This non-detection may be induced by our smaller number of observations compared to K15 and because our GP detectability is significantly degraded by the presence of the additional RV trend induced by the distant stellar companion. When plotting both RV data sets phased along the period from K15 (Fig.~\ref{hd16232ph}), we find that our \sophie~RV and the best model from K15 are not in a perfect agreement; however it is unclear if this is induced by a phase shift between our data and the fit or by the relatively high stellar noise in the data. If combining both RV data sets with {\it yorbit}, we do not achieve a good 1-planet Keplerian fit. Note that we obtain a much better RV accuracy than K15 thanks to the use of \sophie~and \safir. The Lomb-Scargle periodogram of the \sophie~RV is completely dominated by the window function, due to the presence of the RV linear drift. On the contrary, the Lomb-Scargle and CLEAN periodograms of the more consequent K15 RV dataset exhibit a clear peak at the planet period.\\

To investigate the impact of our temporal sampling on HD~16232b detectability, we computed the expected RV given K15 orbital fit at the epochs of our \sophie~observations. To do so, we extrapolated the Keplerian+linear fit from K15 to our observation epochs, and then added RV white noise with the same dispersion as the short-term (1-night) RV jitter we obtained in our \sophie~RV data. These extrapolated RV look the same as our \sophie~RV, the RV linear drift being detectable, but not the planet-induced periodic RV variations (Fig.~\ref{hd16232}, second row). Moreover, the Lomb-Scargle periodogram of the extrapolated RV look very much the same as for our \sophie~RV (\ie~it is dominated by the window function), meaning that the GP detected by K15 is most probably non-detectable here, given our temporal sampling, our low number of observations compared to K15, and the level of short-term RV stellar jitter. Additional \sophie~observations sampled over the orbital period of HD~16232b could allow to raise the ambiguity.\\

In the context of the following statistical analysis, we nonetheless decided to include HD~16232b as a strong candidate GP. Based on the Keplerian fit by K15, we computed HD~16232b minimal mass considering a stellar mass of 1.2 \Msun~to be fully consistent with our stellar mass values, finding a slightly increased \msini~$= 6.8 \pm 1.4$ \Mjup. We include the detailed parameters of the Keplerian fit in Table~\ref{table:orbit_param}.

\renewcommand{\arraystretch}{1.25}
\begin{table*}[t!]
\caption{Stellar properties of our targets with detected GP.}
\label{table:stell_param}
\begin{center}
\begin{tabular}{l c c c}\\
\hline
\hline
Parameter    & Unit  & HD~16232                           & HD~113337                \\              
\hline
Spectral type&       & F6V\tablefootmark{a}               & F6V\tablefootmark{b}     \\  
$V$          &       & 7.09\tablefootmark{c}              & 6.03\tablefootmark{c}    \\   
$B-V$        &       & 0.5\tablefootmark{c}               & 0.37\tablefootmark{c}    \\  
\vsini       & [\kms]& 30\tablefootmark{c}                & 6\tablefootmark{c}       \\    
$\pi$        & [mas] &  $24.52 \pm 0.68$\tablefootmark{d} & $27.11 \pm 0.29$\tablefootmark{d}\\
$[$Fe/H$]$   &       &  -0.03\tablefootmark{e}            &   0.09\tablefootmark{e} \\  
$T_{\rm eff}$  &   [K] & $ 6396 \pm 95$\tablefootmark{e}    & $6669 \pm 80$\tablefootmark{e} \\
$\log{g}$    & [dex] &$4.44 \pm 0.08$\tablefootmark{f}    & $4.21 \pm 0.08$\tablefootmark{f} \\ 
\Mstar       &[\Msun]&$1.20 \pm 0.09$\tablefootmark{f}    & $1.41 \pm 0.09$\tablefootmark{f}\\
Radius       &[\Rsun]&$1.10 \pm 0.05$\tablefootmark{f}    & $1.55 \pm 0.07$\tablefootmark{f} \\
Age          & [Gyr] & $0.9 \pm 0.8$\tablefootmark{a}     & $0.15_{-0.05}^{+0.1}$\tablefootmark{g} \\
\hline
\end{tabular}
\tablefoot{
\tablefoottext{a}{\cite{guenther09}}
\tablefoottext{b}{\cite{boesgaard86}}
\tablefoottext{c}{Taken from the CDS. the uncertainties on the \vsini~are not provided.}
\tablefoottext{d}{\cite{vanleeuwen07}}
\tablefoottext{e}{\cite{casagrande11}. The catalog does not provide the uncertainties on the metallicity.}
\tablefoottext{f}{\cite{allende99}}
\tablefoottext{g}{\cite{borgniet14}}
}
\end{center}
\end{table*}

\renewcommand{\arraystretch}{1}

\renewcommand{\arraystretch}{1.25}
\begin{table*}[t!]
\caption{Best orbital solutions.}
\label{table:orbit_param}
\begin{center}
\begin{tabular}{l l c c c}\\
\hline
\hline
Parameter  & Unit   & HD~16232b~$\dagger$ & HD~113337b & HD~113337c  \\ 
\hline
Status     &        & Probable            & Confirmed  & Possible     \\
\hline
$P$        & day    & $345.4 \pm 3.7$     & $323.4 \pm 0.8$  & $3264.7 \pm 134.3 $ \\          
$T_0$      & BJD-2453000 & $912.8 \pm 42.5$ & $2757.5 \pm 4$ & $5236.8 \pm 217.1 $    \\           
$e$        &             & $0.18 \pm 0.13$  & $0.36 \pm 0.03$ & $0.18 \pm  0.04$      \\        
$\omega$   & \degr       & $-22.9 \pm 44.1$ & $-130.9 \pm 4.9$& $-46 \pm 13 $      \\          
$K$        & \ms         & $176.6 \pm 31.5$ & $76.1 \pm 2.9$  & $76.7 \pm 2.4$     \\         
\hline
Lin.       & \msy        &  $-44.1 \pm 10.2$ &  0             & -                \\
Quad.      & \msysq      &  0                &  0             & -               \\
$N_{\rm m}$  &             & 110 (from \cite{kane15}) & 301                 & -             \\                 
$\sigma_{O-C}$ & \ms      & 128 (228.4)~$^{\star}$  & 22.3 (67.4)~$^{\star}$ & -          \\ 
Reduced $\chi^{2}$ &      & 4.4 (25.4)~$^{\star}$    & 2.9 (8.6)~$^{\star}$  & -          \\     
\hline
\msini     & \Mjup       & $6.8 \pm 1.4$      & $3 \pm 0.3$       & $6.9 \pm 0.6$       \\    
$a_{P}$     & au          & $1.03 \pm 0.01$    & $1.03 \pm 0.02$   & $4.8 \pm 0.23$       \\    
  \hline

\end{tabular}
\end{center}
~$\dagger$ The orbital parameters for HD~16232b have been computed with {\it yorbit} based on the RV measurements given by \cite{kane15} and the stellar mass value from \cite{allende99}. $^{\star}$ The number in parenthesis refers to the model assuming a constant velocity.
\end{table*}
\renewcommand{\arraystretch}{1.}

\subsection{$\theta$ Cyg: a system with complex RV variations}\label{theta_cyg_var}

We detail here our results on HD~185395 or $\theta$ Cyg, a target that we observed extensively along our \sophie~survey and that exhibits intringuingly complex RV and line profile variations. The RV and line profile data of $\theta$ Cyg are illustrated in details in Appendix~\ref{app:details}.

\paragraph{The HD~185395 system --}

We presented our first results on $\theta$ Cyg (F4-5V, 1.37 \Msun) in \cite{desort09}, hereafter D09. We made a first analysis of the RV based on 91 \elodie~spectra and our first 162 \sophie~spectra. We showed in D09 that both \elodie~and \sophie~datasets exhibit a strong quasi-periodic RV signal, with a $\sim$220 \ms~amplitude and a main periodicity of $\sim$150 days. Along with these RV variations was a flat (RV,~BIS) diagram, the BIS variations being of much smaller amplitude ($\sim$50 \ms) than the RV. We argued in D09 that given the 7 \kms~\vsini~of the target, such a flat (RV,~BIS) diagram would usually be a clear evidence that the RV variations are induced by a planetary companion and are not of stellar origin. However, we also showed that despite this flat (RV,~BIS) diagram, the BIS exhibits a strong periodicity of around 150 days, making the origin of this periodicity puzzling. Furthermore, the $\sim$150-day RV variations could not be easily fitted with Keplerian models, with no stable and/or satisfying solution. We concluded in D09 that the origin of these complex RV and BIS variations could not yet be assessed. In addition, we reported the detection by imaging of a wide stellar companion to $\theta$ Cyg, that yet could not be responsible for the $\sim$150-day RV and line profile variations. $\theta$ Cyg was also independently followed in RV by another team \citep{howard16}. Based on 223 RV measurements obtained at the Lick Observatory, these authors also detected the long-term RV drift, as well as the quasi-periodic RV variations at $\sim$150 days, that they deemed statistically non-significant.

Finally, $\theta$ Cyg has been a target of interest in optical and IR interferometry. Multiple \vega~observations allowed \cite{ligi12} to make a first estimation of $\theta$ Cyg angular diameter. Furthermore, the unusually high jitter and a possible periodic variability in the squared visibility measurements led these authors to speculate on the presence of a low-mass, unseen close stellar companion as the possible source of the RV and interferometric jitter. However, the more recent study of \cite{white13}, based notably on IR closure phase (CP) measurements made with the Michigan Infrared Combiner \citep[MIRC,][]{monnier04} interferometer on CHARA, found no evidence for a close stellar companion to $\theta$ Cyg. Based on the CP, these authors derived upper limits on a potential companion brightness, and found that a companion would have to be fainter than the primary by at least 4.7 magnitudes in the $H$-band between 0.2 and 0.4 au, and fainter by at least 3.4 magnitudes between 0.4 and 0.7 au. Using stellar mass-luminosity relations \citep{baraffe98} for an age of $\sim$1 Gyr, this translates into upper masses of 0.3 and 0.5 \Msun, respectively.

\paragraph{New \sophie~data --}

We acquired 164 additional spectra on $\theta$ Cyg from 2009 to 2016, raising the total number of \sophie~spectra to 326, and extending the \sophie~time baseline to 3482 days ($\sim$9.5 years). We display the main \sophie~spectroscopic observables in Fig.~\ref{hd185395}. The new \sophie~RV data set show both a long-term drift of slightly quadratic shape as well as the quasi-periodic $\sim$150-day variations. The quadratic trend is probably induced by the wide stellar companion imaged by D09 (see more details in Sect.~\ref{subsubsect:thetacyg_BIN}). Once corrected from this quadratic trend, the remaining RV variations have a total amplitude of 275 \ms~and a dispersion of 64.5 \ms. The RV Lomb-Scargle periodogram exhibits several peaks above the $1\%$ FAP between 100 and 500 days, but only a single peak at 150 days remains when cleaning the periodogram from the window function (CLEAN, Fig.~\ref{hd185395}).

\object{HD\,185395} BIS shows only low-amplitude variations (rms 12.5 \ms), thus leading to a flat (RV,~BIS) diagram, indicative of a companion. However, the BIS Lomb-Scargle periodogram exhibits a single peak at a 145-day periodicity (in agreement with D09) which is also present in the CLEAN periodogram. The FWHM shows periodic variations with a 207 \ms~total amplitude (rms 38 \ms), corresponding to a single peak at 149 days on the FWHM periodograms. The amount of correlation between the RV corrected from the quadratic fit and the FWHM is low (-0.3), but slightly higher than between the uncorrected RV and the FWHM (-0.2). Both the BIS and FWHM data can be fitted fairly well with a nearly sinusoidal model with {\it yorbit}, with periods of 144 and 148 days, respectively, and marginally compatible phases. Finally, we remind that $\theta$ Cyg does not show any chromospheric emission in the Calcium lines ($<$ \rhk~$>$ = -4.82).

\paragraph{Origin of the RV variations --} As already showed by D09, a purely stellar origin to the $\sim$150-days RV and line profile variations is unlikely. Given $\theta$ Cyg \vsini~of $\sim$7 \kms, the rotational modulation of stellar magnetic activity should induce RV-correlated BIS variability, which is inconsistent with the flat (RV,~BIS) diagram. Stellar granulation is not known to induce such high-amplitude RV variations at such a long period, whereas magnetic activity cycles last much longer given our present knowledge \citep[several years, see][]{baliunas95,lovis11}. Finally, most of stellar pulsations (\ie~solar-like or $\gamma$ Dor-type) happen at much shorter periodicities; {\it Kepler} data have revealed solar-like oscillations on $\theta$ Cyg, but no other clear pulsation pattern \citep{guzik11}.

A second hypothesis could be the presence of a close, unseen low-mass stellar companion, which was not explored by D09. Such a companion would be just bright enough to slightly blend the primary spectrum and induce the $\sim$150-day periodic variation in both the RV and the line profile observables without significantly altering the CCF shape. \cite{santerne15} showed that under some configurations, such an unresolved double-lined spectroscopic binary (hereafter SB2) would produce only a very weak correlation between the RV and the BIS, making it looking like a planetary signal. However, these authors also emphasized that even in such configurations, the FWHM should still be correlated to the RV, while here the $\theta$ Cyg RV-FWHM correlation is weak. In contrast, a more convincing example of a RV-FWHM correlation probably induced by a spectroscopic binary can be found for HD~191195 (Sect.~\ref{subsubsect:hd191195}). For masses between $\sim$0 (\ie, negligible compared to the primary) and 0.5 \Msun, such a companion on a (nearly circular) 150-day orbital period would have a sma between 0.6 and 0.7 au, respectively. This falls into the separation range covered by the interferometric detection limits ($\leq$ 0.5 \Msun) from \cite{white13}. The $H$-band contrasts provided by these authors may well translate into larger contrasts in the $V$-band (\ie, the \sophie~range) that would exclude significant effects on the CCF and RV. While not conclusive, this is another argument against the stellar companion hypothesis.

Finally, the planetary (or multi-planetary) hypothesis was already well explored by D09, but with no satisfying results. We tried to fit various Keplerian models with {\it yorbit} to $\theta$ Cyg RV (from 1 to 4 planets, with/without trends) and considered the cases detailed by D09, but to no avail. Furthermore, we also considered the \elodie~RV data set from D09, and the RV data set published by \cite{howard16}. Both these data sets exhibit RV variations of $\sim$300 \ms~amplitude with a periodic character at about 150 days. We also tried to fit these different RV data sets with {\it yorbit}, either separately or by combining them. However, we did not get any satisfying solution. Furthermore, the main $\sim$150-day RV periodicity seems rather variable on our timebase, going from $\sim$125 days to $\sim$180 days if dividing our timebase in several slices (Appendix~\ref{app:details}).

In conclusion, we consider that the origin of $\theta$ Cyg mid-term RV variations cannot be definitely determined. The variability of the RV variation period, the similarity of the RV, BIS and FWHM periodograms and the flat (RV,~BIS) diagram make it a truly puzzling case.

\subsection{RV long-term trends and stellar binaries}\label{subsect:BIN}

\begin{figure*}[ht!]
  \centering
\includegraphics[width=0.95\hsize]{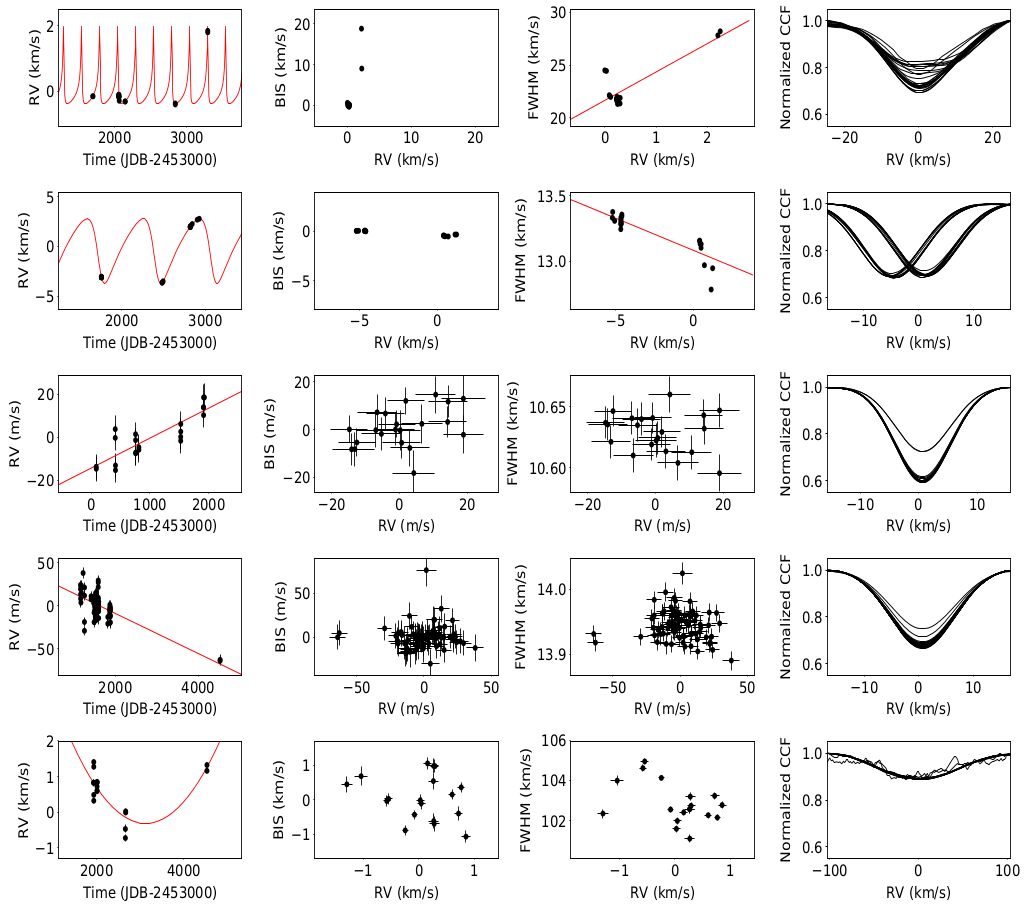}
\caption{Candidate binary companions ({\it first part}). {\it From top to bottom}: HD~10453, HD~20395, HD~43318, HD~82328, HD~143584. From left to right: RV time series, BIS vs RV, FWHM vs RV, stacked CCF. The best {\it yorbit} fit of the RV variations is overplotted in red to the RV. On the RV-FWHM diagram, the RV-FWHM correlation is overplotted in red if the absolute Pearson correlation coefficient is $\geq$ 0.6.}
\label{sb1}
\end{figure*}

In this section, we describe fourteen massive and/or distant companions that we confidently identified in our survey. Given our data, most of them must be spectroscopic binaries (SB), while the others are candidate SB. We proceeded in the same way as done and explained before in Paper IX: 
\begin{enumerate}
\item We identified the companion presence based on various diagnosis -- RV variations, flat (RV,~BIS) diagram, CCF variability, RV-FWHM correlation.
\item We distinguished between double-lined (SB2) and single-lined (SB1) binaries based on the presence (or absence, respectively) of CCF distortions and/or RV-FWHM correlation. 
\item If possible, we fitted the RV with various models (linear, quadratic or keplerian) using both our \sophie~data and, whenever possible, other RV datasets available in the literature. We then explored the RV residuals looking for lower-mass companions.
\item For our targets with long-term (linear or quadratic) RV trends, we constrained the companion properties (\msini, sma) given the available data, as we did in Paper IX. Briefly, we assumed the companion period to be at least equal to the observation time baseline, and the RV amplitude induced by the companion to be at least equal to the span of the observed RV trend, considering a circular orbit. We then deduced the corresponding minimal mass vs sma relation (see Paper IX for more details). If appropriate, we looked for additional constraints from the litterature.
\end{enumerate}

\begin{figure*}[ht!]
\centering
\includegraphics[width=0.7\hsize]{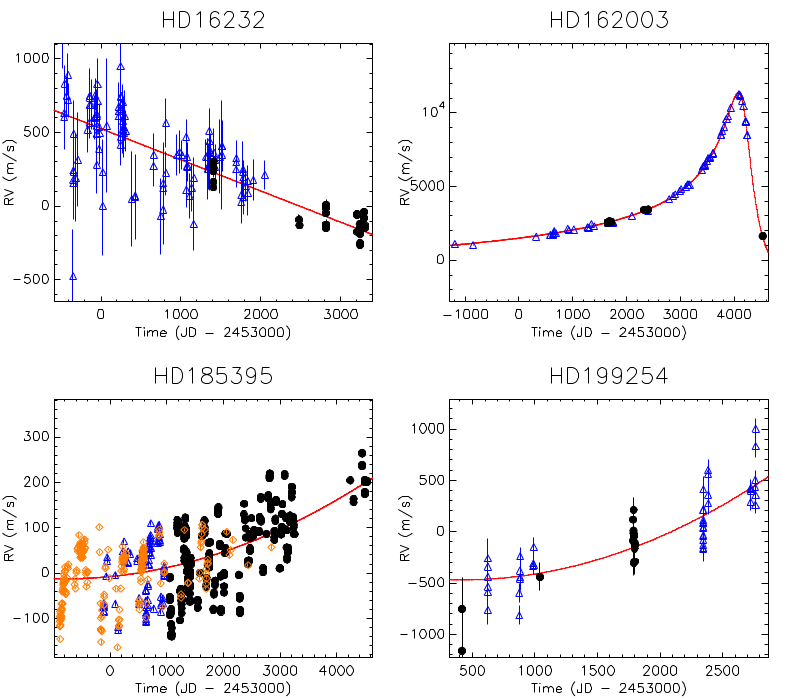}
\caption{Targets with combined RV datasets. On all figures, RV data from our \sophie~survey are displayed as black dots, and the {\it yorbit} fit we
derived from the combined RV data sets (either linear, quadratic, or Keplerian) is overplotted in red. {\it Top left}: HD~16232 -- RV data from \cite{kane15} are displayed as blue triangles. {\it Top right}: HD~162003 -- RV data from \cite{endl16} are displayeded as blue triangles. {\it Bottom left}: HD~185395 -- our \elodie~RV data \citep{desort09} are displayed as blue triangles and RV data from \cite{howard16} are displayed as orange diamonds. {\it Bottom right}: HD~199254 -- our \harps~RV data (Paper IX) are plotted as blue triangles.}
\label{multi}
\end{figure*}

\subsubsection{HD~10453}

HD~10453 (HIP~7916, F5V) shows high-amplitude ($\sim$2.2~\kms~peak-to-peak) RV variations over our 1590-day time baseline, along with a distorted CCF and high-amplitude BIS and FWHM variations (Fig.~\ref{sb1}) that are characteristic of a SB2 binary. We tried to fit the RV variations with {\it yorbit} using a Keplerian model. When letting free all the orbital parameters, the best model corresponds to a companion with \msini~= 31.1~\Mjup, $P = 249.2 \pm 20.5$ days and $e \simeq$ 0.76. However, given {\it i}) the obvious SB2 nature of this companion considering the line profile variations, and {\it ii}) that our temporal sampling is far from completely covering the orbital phase of the binary secondary component, we consider that the actual companion mass has to be much larger than this estimation. Note that if running {\it yorbit} with various forced periods in the range 100-3000 days, the solutions we find are almost as good as the free-parameter one in terms of $\chi^{2}$, highlighting our incomplete temporal sampling. 

HD~10453 is a known binary. A stellar companion was resolved by speckle interferometry at a projected separation of $\sim$0.05'' \citep[$\sim$2 au,][]{hartkopf12} while \cite{riddle15} imaged a stellar companion at a projected separation of $\sim$0.18'' ($\sim$7 au). Given the uncertainties on our orbital parameters, we cannot conclude whether the RV variations are induced by the companion detected by \cite{hartkopf12} or \cite{riddle15}, or whether the HD~10453 system has more components.

\subsubsection{HD~16232}\label{subsubsect:hd16232_BIN}

\begin{figure*}[ht!]
\centering
\includegraphics[width=1\hsize]{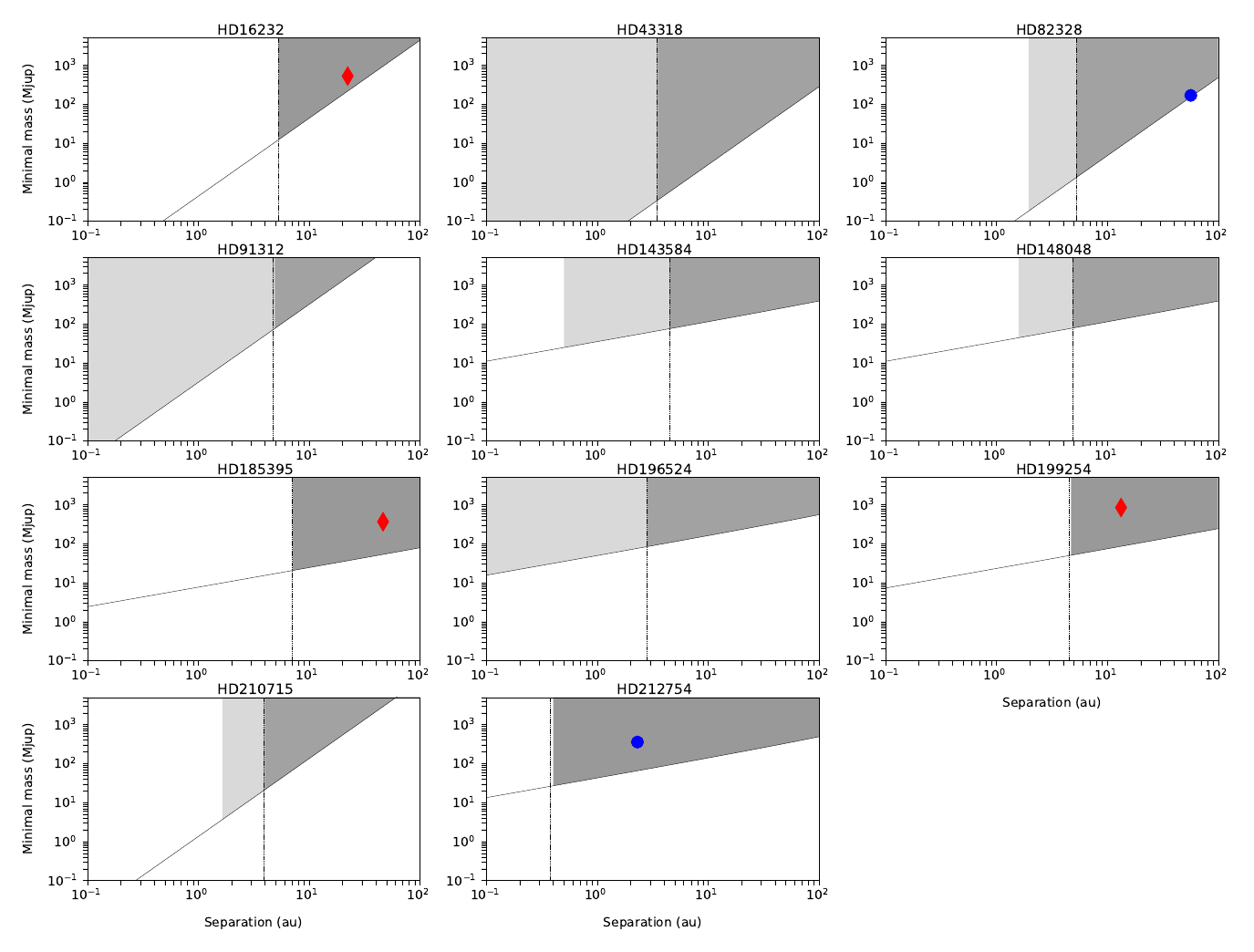}
\caption{Constraints on the \msini~and sma of the companions to our targets with RV linear or quadratic trends. {\it Dark gray}: possible (mass, sma) domain for the companion given the available RV data, assuming a circular orbit and an orbital period at least equal to our total time baseline. {\it Light gray}: extended (mass, sma) domain in the case orbital periods smaller than our total time baseline can still be considered, given the observation temporal sampling. {\it Red diamonds}: previously imaged companions from the literature: HD~16232 -- DI from \cite{roberts15}; HD~185395 --
DI from \cite{desort09}; HD~199254 -- DI from \cite{derosa14}. {\it Blue dots}: companions previously detected with astrometry and/or RV from the literature: HD~82328 -- astrometry from \cite{tokovinin14}; HD~212754 -- RV from \cite{griffin10}, and astrometry from \cite{goldin07} and \cite{tokovinin14}.}
\label{constraint_bin}
\end{figure*}

As presented in Sect.~\ref{subsubsect:hd16232_GP}, we detect a long-term linear trend in our \sophie~RV, along with a composite (RV,~BIS) diagram indicative of a companion-induced trend (Fig.~\ref{hd16232}). A linear RV trend was already reported by \cite{kane15}. To derive the best constraints on this companion, we combined our \sophie~RV data set with the RV from K15 and fitted a linear RV drift with {\it yorbit}. Such a combination of different RV data sets is possible as {\it yorbit} includes the RV offsets between the data sets as fitting parameters. The best solution is a linear trend with a $-77.4 \pm 3.2$ \msy~slope and a 802 \ms~amplitude over the combined RV data set (Fig.~\ref{multi}). The companion responsible for this RV trend has then to be more massive than 10 \Mjup~and has to be orbiting further than 5.2 au (Fig.~\ref{constraint_bin}).

A stellar companion to 30 Ari B has been imaged at a projected separation of $\sim$0.54'' \citep[$\sim$22 au,][K15]{riddle15}, with a mass estimated to be roughly 0.5 \Msun~\citep{roberts15}. In addition, \cite{kane15} estimated that the mass of the imaged companion would have to be higher than 0.29 \Msun~($\sim$304 \Mjup) to explain the RV linear trend they detected. The mass and separation estimated by \cite{roberts15} for the companion they imaged are compatible with our constraints on the minimal mass and sma of the companion responsible for the long-term RV trend detected both in our \sophie~data and in K15 data (Fig.~\ref{constraint_bin}).

\subsubsection{HD~20395}

HD~20395 (14 Eri, HIP~15244, F5V) shows high-amplitude ($\sim$6.4 \kms) RV variations along with a flat (RV,~BIS) diagram, large variations of the CCF, and an apparent RV-FWHM anti-correlation (Fig.~\ref{sb1}) that are characteristic of a spectroscopic binary. We used the {\it yorbit} software to fit the RV variations. The best solution corresponds to a low-mass stellar companion with $P = 675.2 \pm 5.9$ days, \msini~$\sim$163 \Mjup~and $e = 0.33 \pm 0.02$ (corresponding to a $\sim$1.7 au sma). However, given that our orbital phase coverage is not complete, we cannot exclude longer periods (and larger \msini) for HD~20395B. If constraining the companion orbital period to larger ranges (up to $2.10^{4} - 3.10^{4}$ days) with {\it yorbit}, reliable orbital solutions can be found (with sma up to $\sim$15 au, \msini~up to $\sim$350 \Mjup, and eccentricities up to 0.8). HD~20395 has been reported as an astrometric binary based on its proper motion \citep{makarov05,frankowski07}, but without estimations of the orbital parameters.

\subsubsection{HD~43318}

HD~43318 (HIP~29716, F5V) shows a small ($5 \pm 0.7$ \msy) long-term linear drift in its RV over our $\sim$1800-day timebase, along with a possibly composite (RV,~BIS) diagram (Fig.~\ref{sb1}). We thus classify this trend as a candidate companion-induced one. When corrected from the linear trend, the RV residuals show a dispersion two times smaller than the RV, and the Lomb-Scargle periodogram of the residuals show significantly less power at periods $>$ 100 days. Given the small amplitude of the RV trend over our timebase, the candidate distant companion can be either of GP, BD or stellar nature (Fig.~\ref{constraint_bin}). HD~43318 has not been reported as a binary before.

\subsubsection{HD~82328}

\begin{figure}[ht!]
\centering
\includegraphics[width=0.82\hsize]{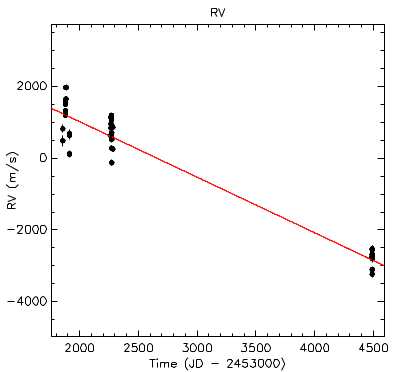}
\caption{HD~91312 RV time series. The assumed linear fit is overplotted in red to the RV.}
\label{hd91312}
\end{figure}

$\theta$ UMA (HIP~46853, F7V) shows a linear RV drift with a 78 \ms~amplitude over our 3368-day timebase, that is best fitted with a long-term linear trend (Fig.~\ref{sb1}). If assuming that the period of the companion responsible for the RV trend is at least equal to the total observation timebase, the companion can then be either of BD or stellar nature and has to be orbiting further than 5.3 au from the primary (Fig.~\ref{constraint_bin}). HD~82328 is a known physical binary with a $\sim$5'' projected separation \citep{allen00}. Recently, \cite{tokovinin14} has estimated the orbital parameters of the companion based on the absolute $V$-magnitudes of the components and assuming that the projected separation corresponds to the actual sma of the companion, finding $P \sim$322 years and \msini~= 0.16 \Msun. Such a companion is compatible with the detected RV trend (Fig.~\ref{constraint_bin}).

\subsubsection{HD~91312}

HD~91312 (HIP~51658, A7IV) exhibits a large RV drift (of $\sim$4.1 \kms~amplitude) over our 2635-day timebase (Fig.~\ref{hd91312}). As we cannot compute the line profiles for this target (too few spectral lines are available for \safir~computation), we cannot check the (RV,~BIS) diagram. Yet, we conclude that this RV drift is most probably induced by a companion because its amplitude is far larger than the short-term RV dispersion induced by stellar pulsations for this spectral type. This long-term RV drift is best fitted by a linear trend with a $-566 \pm 25$ \msy~slope; once corrected from the drift, the RV dispersion decreases from 1.75 \kms~to 415 \ms. 

HD~91312 is a known wide (23'' or $\sim$796 au) visual binary, which is dynamically linked according to \cite{kiyaeva08}. The very large projected separation of this companion makes it unlikely to be at the origin of the RV drift we detect. This RV trend is then likely induced by an unknown companion. Assuming an orbital period larger than our time baseline, this companion has then to be of stellar mass and has to orbit further than 4.7 au around the primary (Fig.~\ref{constraint_bin}). Interestingly, HD~91312 is still young ($\sim$200 Myr) and shows an IR excess characteristic of the presence of a cold debris disk, according to \citep{rhee07}. A recent SED analysis \citep{rodriguez12} gives a dust radius of 179 au; this could then be a circumbinary disk.

\subsubsection{HD~143584}

HD~143584 (HIP~78296, F0IV) exhibits high-amplitude ($\sim$2 \kms) RV variations over our 2596-day timebase on both short and long terms, along with a composite (RV,~BIS) diagram (Fig.~\ref{sb1}). This is best explained by the presence of both a massive companion and stellar pulsations. The RV variations are best fitted with a long-term quadratic trend, even if periods shorter than our time baseline could still be possible. The dispersion of the RV residuals is 291 \ms, compared to the 592 \ms~RV dispersion. Assuming an orbital period larger than our time baseline, the companion responsible for the RV long-term variations would have to be of stellar nature and would have to orbit further than 4.5 au from the primary (Fig.~\ref{constraint_bin}). HD~143584 has not been reported as a binary before.

\subsubsection{HD~148048}

\begin{figure*}[ht!]
\centering
\includegraphics[width=0.95\hsize]{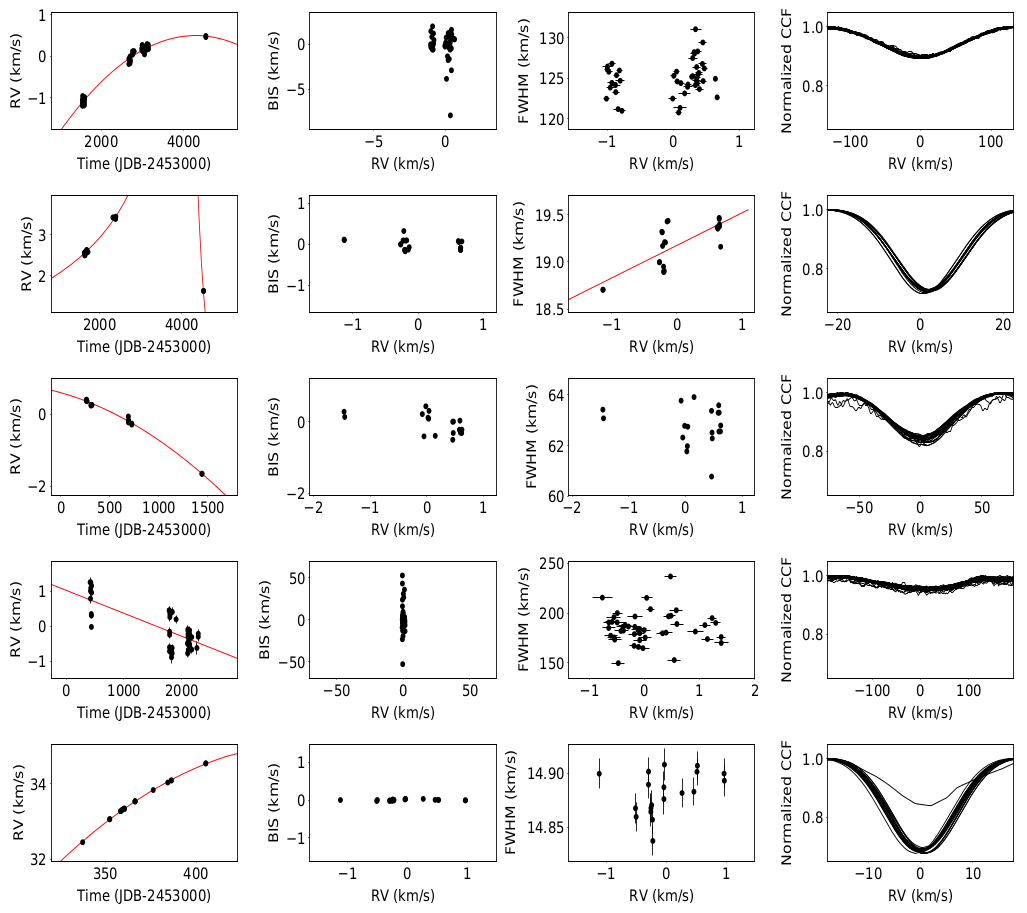}
\caption{Candidate binary companions ({\it second part}). From top to bottom: HD~148048, HD~162003, HD~196524, HD~210715, HD~212754. {\it From left to right}: RV time series, BIS vs RV, FWHM vs RV, stacked CCF. The best {\it yorbit} fit of the RV variations is overplotted in red to the RV. On the RV-FWHM diagram, the RV-FWHM correlation is overplotted in red if the absolute Pearson correlation coefficient is $\geq$ 0.6.
}
\label{sb2}
\end{figure*}

$\eta$ Umi (HIP~79822, F5V) exhibits long-term RV variations that are best fitted by a quadratic trend of 1600 \ms~amplitude over our 2988-day timebase. The (RV,~BIS) diagram is composite with an horizontal spread induced by a companion and a vertical spread induced by stellar pulsations (Fig.~\ref{sb2}). Assuming an orbital period larger than our time baseline, the companion responsible for these RV variations would have to be of stellar nature and would have to orbit further than 4.8 au from the primary (Fig.~\ref{constraint_bin}). HD~148048 has not been reported as a binary before.

\subsubsection{HD~162003}

$\psi^{1}$ Dra A (HIP~86614, F5IV-V) shows high-amplitude ($\sim$2~\kms), long-term RV variations over our 2890-day time baseline, along with a flat (RV,~BIS) diagram (Fig.~\ref{sb2}) characteristic of a companion. Furthermore, the RV and FWHM variations are strongly correlated (Pearson's coefficient of 0.75), which make it a slightly SB2 binary.

HD~162003 has previously been reported as a spectroscopic binary. Based on $\sim$40 RV measurements spanned over $\sim$1550 days, \cite{toyota09} detected a long-term RV trend of quadratic shape that they attributed to an unseen companion of \msini~$\sim$50 \Mjup, with a sma lower than 140 au (assuming a circular orbit) to remain stable given the existence of the wide binary $\psi^{1}$ Dra B (HD~162004) at a projected separation of $\sim$30'' ($\sim$667 au) from HD~162003. \cite{gullikson15} detected this unseen companion (hereafter $\psi^{1}$ Dra C) by looking for a secondary peak in the CCF of HD~162003 spectra, in a way similar to \cite{bouchy16}. From this analysis, they estimated that $\psi^{1}$ Dra C had a mass of $\sim$0.7 \Msun, making it of stellar nature, and a sma of $\sim$9 au. Finally, \cite{endl16} directly detected $\psi^{1}$ Dra C with speckle imaging on the one hand, and, based both on 85 RV measurements spanning 15 years (2000-2015) on the other hand, estimated the following orbital parameters: \msini~$\sim 550 \pm 5$ \Mjup, $e = 0.67$, $P = 6650 \pm 160$ days, corresponding to a $\sim 8.7 \pm 0.1$ au sma. We combined our \sophie~data with the RV from \cite{endl16} and fitted a single Keplerian model to the combined RV data set with {\it yorbit}. Our best solution corresponds to a companion with a $661 \pm 107$ \Mjup~minimal mass, a sma of $24.3 \pm 3.8$ au and an eccentricity $e = 0.87 \pm 0.02$ (Fig.~\ref{multi}). We conclude that the RV variations that we detect in our \sophie~data are induced by $\psi^{1}$ Dra C. However, the differences between our Keplerian model and the one derived by \cite{endl16} show that the sampling of the companion orbit is still not complete enough to adequately cover the period, meaning that large uncertainties still remain on its orbital parameters.

\subsubsection{HD~185395}\label{subsubsect:thetacyg_BIN}

As presented in Sect.~\ref{theta_cyg_var}, we detect a long-term quadratic RV drift in addition to the RV mid-term complex variability of $\theta$~Cyg. This specific RV long-term trend is induced by a distant companion, as showed by the flat (RV,~BIS) diagram (Fig.~\ref{hd185395}). To derive the best constraints on this companion, we combined our \sophie~RV data set to the RV acquired with \elodie~before \citep{desort09} and to the RV data from \cite{howard16}. This allows to expand the timebase to 5422~days (14.8~years). Over the combined RV data set, the RV quadratic trend induced by the distant companion has an amplitude of $\sim$370~\ms~(Fig.~\ref{multi}). The companion responsible for this trend is then either of BD or stellar nature and has to orbit further than 7 au from the primary (Fig.~\ref{constraint_bin}).

A distant stellar companion to $\theta$ Cyg was imaged by \cite{desort09} as a projected separation of 46.5 au from the primary. Based on the measured contrast between the two components and on stellar evolutionary models, these authors deduced a mass of $\sim$0.35 \Msun~for the companion. These estimated mass and projected separation are compatible with the constrains we derive from the RV long-term trend (Fig.~\ref{constraint_bin}). 

\subsubsection{HD~191195}\label{subsubsect:hd191195}

HD~191195 (F5V, 1.49 \Msun) exhibits complex RV variations, with a RV peak-to-peak amplitude of 272 \ms~and a dispersion of 56.5 \ms~(Fig.~\ref{hd191195}). We consequently followed this target intensively, acquiring 265 spectra over a 3191-day ($\sim$8.7 years) time baseline. The RV Lomb-Scargle shows multiple peaks between 100 and 10$^{3}$ days, but they are all aliases of a single $\sim$300-day periodicity, as showed by the RV CLEAN periodogram. The BIS does not show high-amplitude variations (dispersion of 18 \ms), hence the flat (RV,~BIS) diagram. As the RV, the FWHM exhibits high-amplitude (266 \ms~peak-to-peak) complex variations. The FWHM Lomb-Scargle and CLEAN periodograms are strikingly similar to the RV ones, with a clear, single peak at $\sim$300 days in the CLEAN periodogram. We emphasize that HD~191195 RV and FWHM are strongly anti-correlated, with a Pearson’s correlation coefficient of -0.65 (Fig.~\ref{hd191195}).

Given HD~191195 \vsini~of 5.5 \kms, the flat (RV,~BIS) diagram points towards a companion as the source of the observed RV variability. We tried to fit single or multi-companion Keplerian models to HD~191195 RV, including one to three planets and an additional linear or quadratic trend. However, none of them is able to reproduce convincingly the RV variations. We consider that this complex RV variability is induced by a faint stellar companion rather than by a planetary companion. This is strongly supported by the observed RV-FWHM anti-correlation, which can be explained by an unresolved SB2 binary \citep[see][and the discussion about $\theta$ Cyg in Sect.~\ref{theta_cyg_var}]{santerne15}. Given HD~191195 relatively small \vsini, we might be seeing this possible binary system under an inclined or even close to pole-on configuration, which would explain such a low RV amplitude for a companion of stellar nature. HD~191195 has not been reported as a binary before.

\subsubsection{HD~196524}

$\beta$ Del (HIP~101769, F5IV) shows high-amplitude (2 \kms) RV variations over our 1179-day timebase, along with a flat (RV,~BIS) diagram characteristic of a companion (Fig.~\ref{sb2}). The RV are best fitted by a quadratic model. Assuming an orbital period larger than our time baseline, the companion responsible for this RV trend would have to be of stellar mass and would have to orbit further than 2.8 au from the primary (Fig.~\ref{constraint_bin}).

\begin{figure*}[ht!]
\sidecaption
\includegraphics[width=12cm]{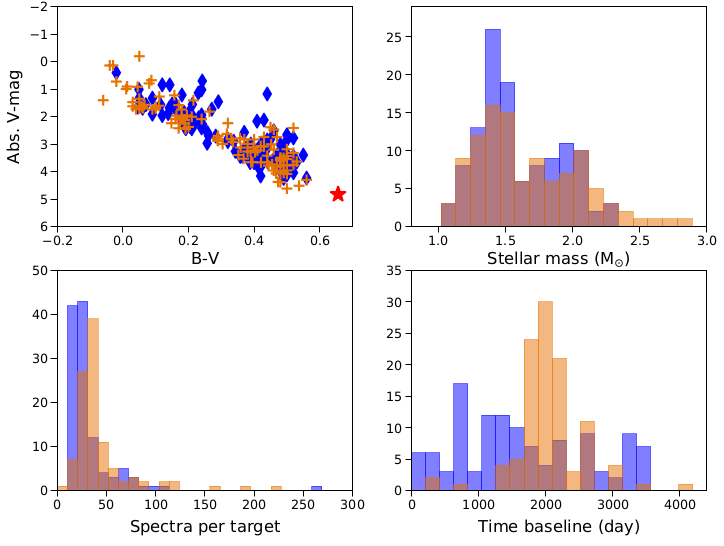}
\caption{Properties of our combined sample (\sophie~+\harps). {\it Top left}: position of our targets in an H-R diagram. {\it Blue diamonds}: \sophie~targets; {\it orange crosses}: \harps~targets. The Sun is displayed ({\it red star}) for comparison. {\it Top right}: stellar mass histogram of our combined sample. {\it Blue}: \sophie~targets; {\it orange}: \harps~targets. {\it Bottom left}: spectrum number per target. {\it Bottom right}: timebase histogram.}
\label{combined_sample}
\end{figure*}

HD~196524 was already known as a spectroscopic binary \citep{pourbaix04}. Based on astrometric data, \cite{malkov12} estimated its period to be of $\sim$26.7 years, its eccentricity of 0.36 and its sma of $\sim$0.44'' ($\sim$13.6 au). At such a separation and given the mass constraints we bring, such a companion would have a mass larger than $\sim$200 \Mjup~(Fig.~\ref{constraint_bin}).

\subsubsection{HD~199254}

We already detected HD~199254 (HIP~103298, A5V) as a spectroscopic binary thanks to our \harps~RV (see Paper IX). We also observed this target with \sophie~and we detected a long-term, high-amplitude RV drift. When combining the \sophie~and \harps~RV data and fitting a trend model with {\it yorbit}, we found the best solution to be a slightly quadratic trend of 923 \ms~amplitude over the total time baseline of 2400 days (Fig.~\ref{multi}). Given our combined RV data, the companion responsible for such a RV trend is of stellar nature and orbits further than 4.5 au from the primary (Fig.~\ref{constraint_bin}). A stellar companion with properties compatible with these constraints was imaged around HD~199254 by \cite[][see details in Paper IX]{derosa14}. 

\subsubsection{HD~210715}

HD~210715 (HIP~109521, A5V) exhibits a clear RV linear drift of 1.2 \kms~amplitude over the 1880-day timebase. Even if the (RV,~BIS) diagram is dominated by a vertical spread induced by stellar pulsations characteristic of its stellar type (Fig.~\ref{sb2}), we consider that the RV long-term trend is induced by a distant massive companion.

HD~210715 has been reported as an astrometric binary \citep[][]{makarov05,frankowski07}, but no constraints on the companion parameters are available in the literature. Assuming an orbital period larger than our time baseline, the unseen companion responsible for the RV long-term trend would have to be either of BD or stellar nature and would have to orbit further than 3.9 au from the primary (Fig.~\ref{constraint_bin}).

\subsubsection{HD~212754}

We acquired only 19 RV measurements over 68 days on HD~212754 (HIP~110785, F7V), but they are enough to detect a high-amplitude (2 \kms) RV drift of quadratic shape, along with a flat (RV,~BIS) diagram and large CCF variations that are indicative of a spectroscopic binary (Fig.~\ref{sb2}). Given our small timebase, we are not able to fully determine the orbital parameters of the companion. Yet this companion has to orbit further than 0.4 au from the primary and is most probably of stellar nature (Fig.~\ref{constraint_bin}), given the CCF variations.

HD~212754 is a known SB1 \citep{griffin10} and astrometric binary \citep{goldin07} with an orbital period of $\sim$931 days, an eccentricity of 0.3-0.4 and a secondary minimal mass of 0.34 \Msun~\citep{tokovinin14} at a separation of $\sim$59 mas ($\sim$2.3 au). This already known stellar companion is most likely at the origin of the detected \sophie~RV variations.

\section{Combined analysis of the \sophie~+ \harps~surveys}\label{sect:analysis}

Here we combine our \sophie~AF survey with the similar \harps~survey described in Paper IX to make a global analysis in terms of achieved companion detections, sensitivity and statistics in the 1 to 1000-day period range. In addition to the two GP detected in this period range in this \sophie~survey (HD~113337b and HD~16232b), we include the three confirmed GP detected in our \harps~survey around two F6V targets (HD~60532b,~c and HD~111998b) in the following analysis. 

\subsection{Characteristics of the combined survey}

We combine the 125 AF dwarf stars of our \sophie~sample to the 109 targets of our \harps~survey (Paper IX). Note that nine of our \sophie~targets were also part of our \harps~sample: HD~13555, HD~25490, HD~29488, HD~102647, HD~197890, HD~199254, HD~211976, HD~218396 and HD~222368. Our combined AF sample is thus made of $N = 225$ distinct targets. In the case of the targets with both \harps~and \sophie~observations, we selected the instrument for which we had the more RV data and the longest time span (\ie~we used the \sophie~data for HD~102647 and HD~218396, and the \harps~data for the seven other targets) and used the corresponding data in the following. 

We do not combine \harps~and \sophie~RV data for these targets as it is not possible here to compute an accurate value of the zero-point between the two RV data sets, the \safir~RV being relative RV. Furthermore, our detection limits are computed based on an analysis of the RV periodogram (Sect.~\ref{subsect:limdets}). In this context, a \harps~+ \sophie~RV combination might lead to a biased combined periodogram and thus biased detection limits. Hence, using only the RV data set with the most data and the longest time span ensures an unbiased periodogram, but it will lead to slightly more conservative detection limits. Our analysis of the combined survey is then based on 107 targets observed with \harps~and 118 targets observed with \sophie. The combination of these two samples is possible because we use the same target selection process (Appendix~\ref{app:select}). Both samples are thus very similar in terms of stellar physical properties (Fig.~\ref{combined_sample}).

The median time baseline of our \sophie~survey is 1448 days (the mean time baseline being 1640 days), with a median spectrum number per target of 23 (36 in average) acquired during a median number of 11 visits (17 in average). These values are smaller than those obtained for our \harps~sample (see Paper IX and Fig.~\ref{combined_sample}). This difference is mostly explained by the increased performances and better observing conditions of \harps~compared to \sophie. When combining the two surveys, we obtain a median timebase of 1888 days (1832 in average) and a median number of 30 spectra acquired per target.

\subsection{Stellar intrinsic variability}\label{subsect:var}

\begin{figure*}[ht!]
  \centering
\includegraphics[width=1\hsize]{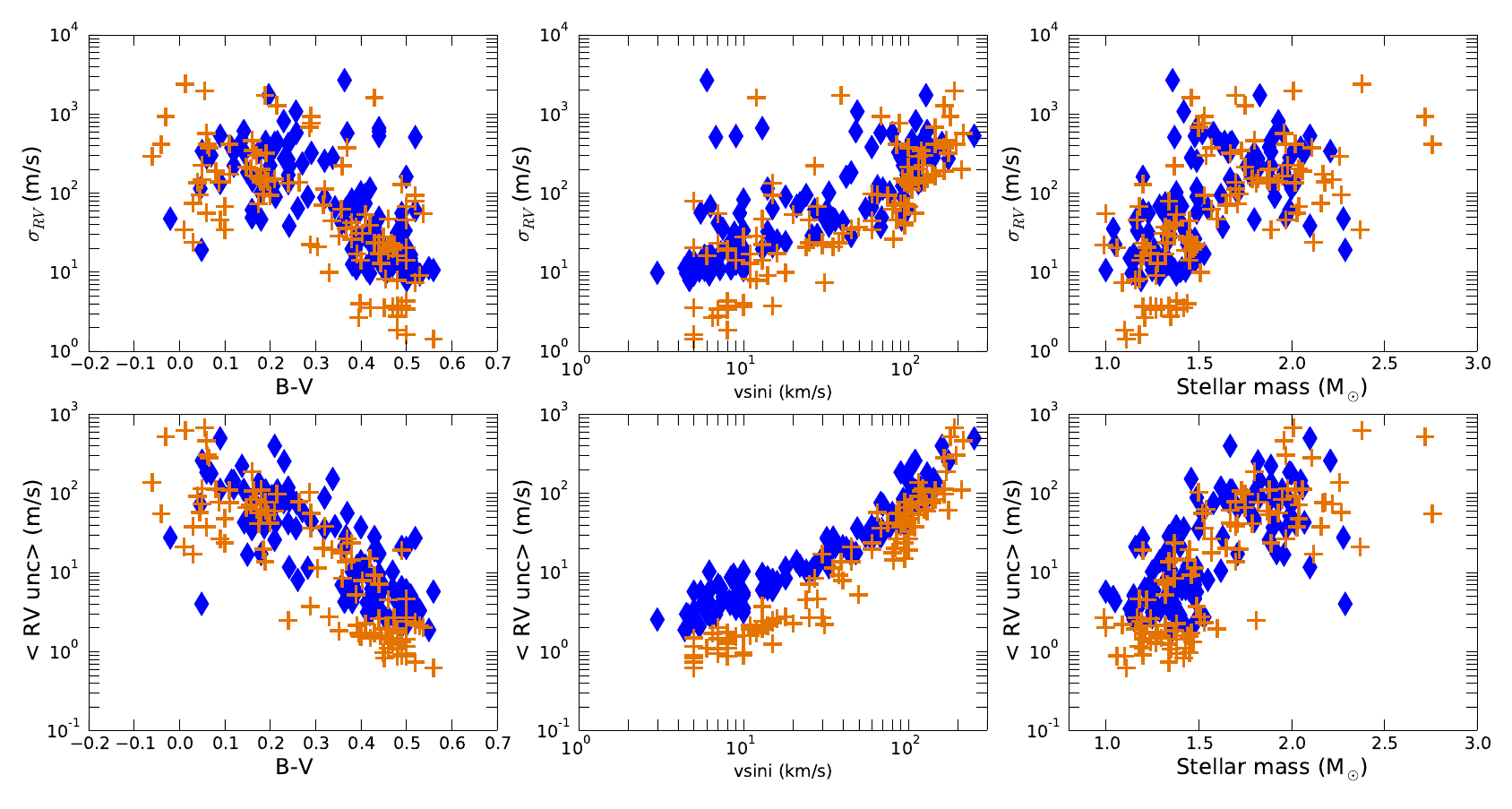}
\caption{Stellar intrinsic RV variability vs stellar properties. {\it First row}: stellar intrinsic RV jitter vs \bv, vs \vsini~and vs mass ({\it from left to right}). {\it Blue diamonds}: \sophie~targets; {\it orange crosses}: \harps~targets. {\it Second row}: averaged RV uncertainty (accounting for the photon noise only) vs \bv, vs \vsini~and vs mass.}
       \label{stell_var}
\end{figure*}

\begin{figure*}[ht!]
  \centering
\includegraphics[width=1\hsize]{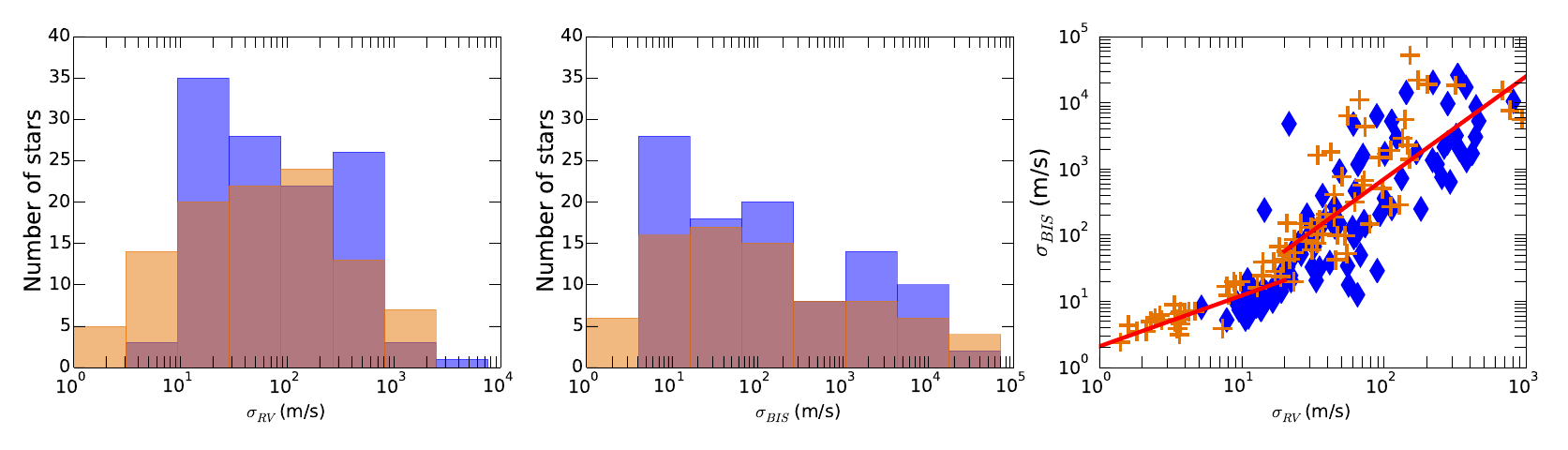}
\caption{Stellar intrinsic variability in our observables. {\it Left}: histogram of the stellar intrinsic RV jitter ({\it blue}: \sophie~targets; {\it orange}: \harps~targets). {\it Middle}: histogram of the BIS dispersion. {\it Right}: BIS dispersion vs intrinsic RV jitter (\sophie~targets are displayed as blue diamonds and \harps~targets as orange crosses). The best power law fits to the BIS rms vs RV rms distribution in the two variability regimes are displayed as red solid lines.}
       \label{hist_var}
\end{figure*}

To characterize the stellar intrinsic variability of our targets, we display in Fig.~\ref{stell_var} the RV dispersion (after having removed the companion-induced RV variations) and the mean RV uncertainties (accounting only for the photon noise) of our targets versus their main physical properties (\bv, \vsini, \Mstar), as we did in Paper IX. Our results are similar to what we found in Paper IX: the RV uncertainties, and, at a somewhat less degree, the RV dispersion, are correlated with the \vsini~and anti-correlated with the \bv, and consequently (loosely) correlated with the stellar mass. In terms of RV and BIS dispersion, the \sophie~and \harps~targets show the same behavior (Fig.~\ref{hist_var}). The only remarkable differences between the two surveys is the larger mean \sophie~RV uncertainty compared to \harps~(Fig.~\ref{stell_var}), and  the relative lack of small RV and BIS dispersion values ($<$~10~\ms) for the \sophie~survey compared to \harps~(Fig.~\ref{hist_var}). This is not induced by a difference between the two stellar samples but rather by the decreased spectral resolution (and thus decreased RV accuracy) of \sophie~with respect to \harps.\\

The median RV dispersion is 61 \ms~for the combined sample (191~\ms~in average), and the median RV uncertainty (not accounting for the instrumental error) is 14 \ms~(49 \ms~in average). As we detailed before in \cite{lagrange09} and Paper IX, the stellar jitter of AF-type MS dwarfs is mainly induced: {\it i}) by stellar magnetic activity (spots and faculae) for our later-type (F) targets, and {\it ii}) by high-frequency pulsations for our earlier-type (A) targets. These two variability regimes are easily seen on a BIS dispersion versus RV dispersion plot (Fig.~\ref{hist_var}). For our targets with a BIS dispersion below $\sim$20 \ms, there is a correlation of 0.7 (Pearson) between the logarithms of the RV and BIS dispersions, and the BIS dispersion increases with the RV dispersion as follows:
\begin{equation}\label{first_eq}
\sigma_{\rm BIS} = 2.1 \ (\sigma_{\rm RV})^{0.78 \pm 0.10}
\end{equation}
In contrast, for our targets with a BIS dispersion above $\sim$20 \ms~(mostly pulsation-dominated jitter), the correlation is 0.73 and the BIS dispersion increases with the RV dispersion as follows:
\begin{equation}\label{second_eq}
\sigma_{\rm BIS} = 0.5 \ (\sigma_{\rm RV})^{1.56 \pm 0.14}
\end{equation}

where the exponent to $\sigma_{\rm RV}$ happens to be twice the one in Eq.~\ref{first_eq}. For a few of our \sophie~targets with a RV jitter dominated by stellar activity, we partially corrected the RV stellar variability by using the RV-BIS anti-correlation, as we did in Paper IX. We used the same criterion, \ie~we made a linear fit of the BIS versus RV data and removed it from the RV if the corresponding Pearson's coefficient was above 0.7 in absolute. This was the case for six of our \sophie~targets (Appendix~\ref{app:sample}). In average, this RV-BIS correction decreases the RV variation amplitude by a 1.6 factor and the RV dispersion (rms) by a 1.7 factor, similarly to what we obtained in Paper IX.

\begin{figure*}[ht!]
  \centering
\includegraphics[width=1\hsize]{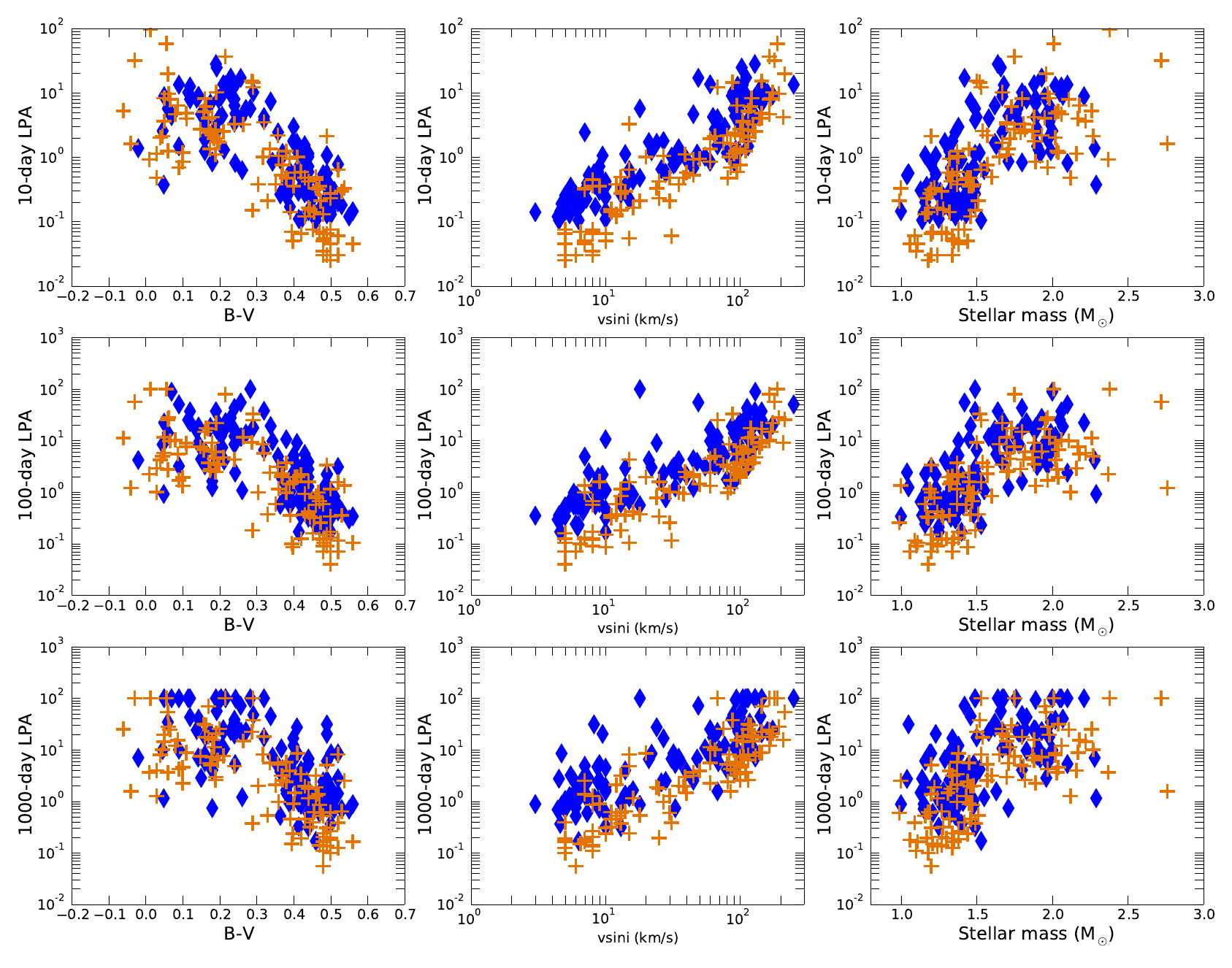}
\caption{Detection limits versus stellar properties. {\it From top to bottom}: 10-day, $10^{2}$-day and $10^{3}$-day LPA detection limits; vs ({\it from left to right}) \bv, \vsini~and mass. \sophie~targets are displayed as blue diamonds and \harps~targets as orange crosses.}
       \label{limdet}
\end{figure*}

\begin{figure*}[ht!]
  \centering
\includegraphics[width=1\hsize]{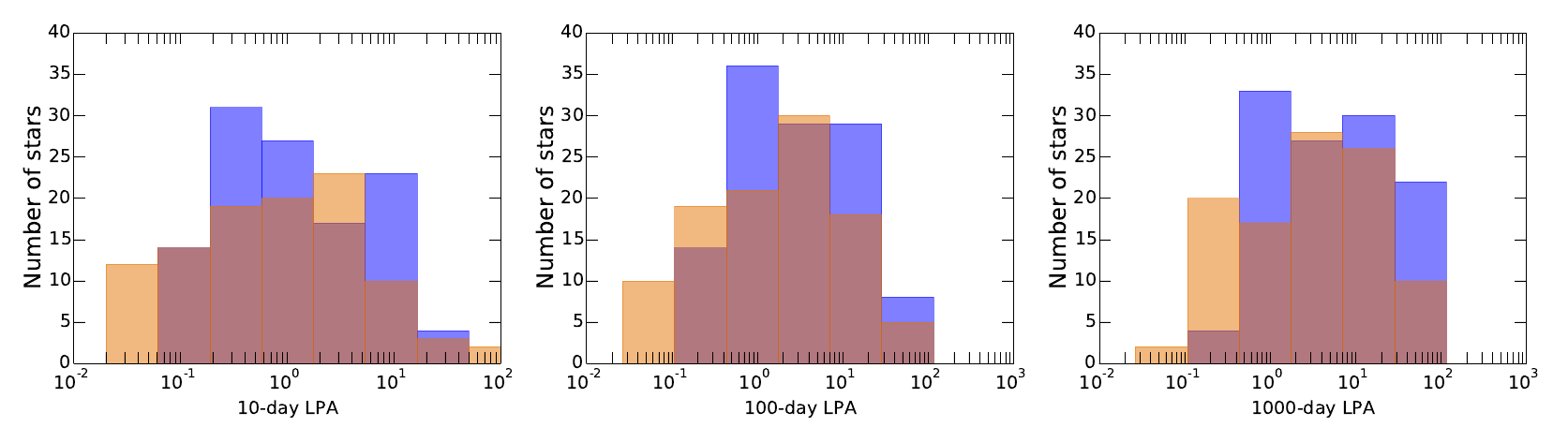}
\caption{Distribution of the achieved detection limits. {\it From left to right}: histograms of the 10-day, $10^{2}$-day and $10^{3}$-day LPA detection limits. {\it Blue}: \sophie~targets; {\it orange}: \harps~targets.}
       \label{limdet_hist}
\end{figure*}

\subsection{Detection limits}\label{subsect:limdets}

We compute the minimal mass (\msini) versus period ($P$) detection limits for each target with no confirmed GP in our combined survey, except for two targets with too few data points ($<$10), for which our periodogram-based detection limits are not effective: HD~196724 and HD~216627 from the \harps~sample (see Paper IX), \ie~219 targets in total. We use the Local Power Analysis (LPA) method \citep{meunier12} to compute the detection limits in the 1 to $2.10^{3}$-day range on a log-spaced grid of 210 periods. Briefly, the LPA method compares the maximum power of the Lomb-Scargle periodogram of a synthetic planet on a circular orbit to the maximum power of the periodogram of the observed RV within a localized period range to compute the \msini~detection limits at each period. Our method as well as the LPA process are fully described in Paper IX.\\

We display our detection limits at meaningful ranges (10, $10^{2}$ and $10^{3}$ days) versus the main stellar properties (\bv, \vsini, mass) in Fig.~\ref{limdet}. On the whole, our results are the same as in Paper IX, \ie~the detection limits are correlated with the \vsini~and anti-correlated with the \bv, in agreement with our results on the intrinsic variability. Overall, the \sophie~targets show slightly higher detection limits than the \harps~targets (Fig.~\ref{limdet_hist}), which is again in agreement with the slightly decreased RV accuracy of \sophie~compared to \harps.\\

\begin{figure}[ht!]
 \centering
\includegraphics[width=1\hsize]{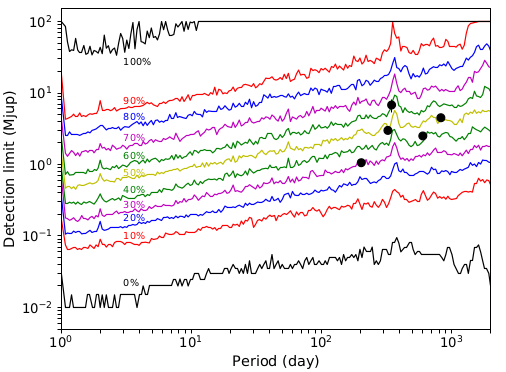}
\caption{Search completeness of our combined survey. {\it From bottom to top}: 0 to 100\%~search completeness (\ie~the fraction of stars with good enough detection limits to rule out a companion of a given \msini~at a given orbital period $P$). {\it Black dots}: confirmed GP detected in the combined survey. Note that HD~16232b (which detection was confirmed by \cite{kane15}) is included here (see Sect.~\ref{subsubsect:hd16232_GP}).}
\label{statlim}
\end{figure}

We compute the search completeness $C$($P$,~\msini) of our combined survey by combining the detection limits of all our targets, as we did in Paper IX for the \harps~targets only. For a given couple ($P$,~\msini), $C$ gives the fraction of stars in the combined sample for which a companion with a minimal mass \msini~and an orbital period $P$ (on a circular orbit) would be detected if present, given our observations. We refer to Paper IX for more details. We display our combined survey completeness in Fig.~\ref{statlim}, along with the confirmed GP detected in the combined survey.

\subsection{Companion occurrence rates}\label{subsect:rate}

\subsubsection{Method summary}

We compute the companion frequency around AF, MS stars based on the detected GP and on the detection limits of our 225-target combined sample. Our method is fully detailed in Paper IX and in Appendix~\ref{app:occ_rate}. Briefly:
\begin{enumerate}
\item We consider different \msini~ranges (0.3 to 1-\Mjup~for ``Saturn-mass'' GP, 1 to 13-\Mjup~for ``Jupiter-mass'' GP, 13 to 80-\Mjup~for BD) and different period ranges (1-10, 10-$10^{2}$ and $10^{2}$-$10^{3}$ days) to define the (\msini, $P$) domains ($D$) where to compute the companion frequency.
\item We consider two target subsamples depending on the stellar mass: first, our 104 targets with masses above 1.5 \Msun, and second our 121 targets with masses $\leq$ 1.5 \Msun.
\item We sum the search completeness $C$($P$, \msini) over \msini~and over $P$ in each defined domain $D$ to get the search completeness function $C_{\rm D}$ over $D$. Excluding the four targets with detected GP (HD~16232, HD~113337 for the \sophie~survey; HD~60532 and HD~111998 from the \harps~survey) and two targets with too few spectra from the \harps~survey (see above), we compute the search completeness for each domain over 219 targets (respectively 102 and 117 targets for the high and low stellar mass subsamples).
\item For each ($P$, \msini) domain, we estimate the number of systems with missed companions $n_{\rm miss}$ that we potentially did not detect in our survey, based on the search completeness $C_{\rm D}$ and the number $n_{\rm det}$ of systems with one (or more) detected GP.
\item We derive the companion occurrence rate in each domain and for our different mass subsamples using binomial statistics, \ie~by computing the probability distribution function (PDF) of drawing $n_{\rm det}$ systems with detected GP among $N = 225$~stars, and then compute the product of this PDF by ($n_{\rm det} + n_{\rm miss}$)/$n_{\rm det}$ to correct for our search incompleteness (see details in Paper IX). The companion occurrence rate corresponds to the companion frequency value with the maximum probability (\ie~to the mode of the PDF). If $n_{\rm det} = 0$, we consider $n_{\rm det}^{'} = 1$ (as an upper limit) and compute the corresponding upper limit on the occurrence rate. 
\end{enumerate} 

\subsubsection{Results from the combined \sophie~+\harps~survey}

We display our results in details in Table~\ref{tab_occur} and in Fig.~\ref{occur}.

\paragraph{Companion frequency vs \msini}
\begin{enumerate}
\item We first consider companions with masses in the 13-80 \Mjup~range. We do not detect any such companion in the 1-$10^{3}$ day range. Our completeness in this \msini~domain is close to 100\%. We thus obtain an upper limit on the BD occurrence rate of $\lesssim$ 3.5-4\%~(at a 1$\sigma$ uncertainty, as we will write hereafter) for both our A- and F-type samples for periods below 1000 days. The differences in occurrence rates between our stellar mass subsamples or between the different considered period ranges are negligible (Fig.~\ref{occur}). Given that these estimations are still upper limits on the occurrence rates (no actual detection), they are compatible with our previous statistical analysis on \harps~targets only (Paper IX). These upper limits on the BD-mass companion occurrence rates are roughly twice smaller than in Paper IX (which is in agreement with a sample twice larger).

\begin{figure}[ht!]
\centering
\includegraphics[width=0.95\hsize]{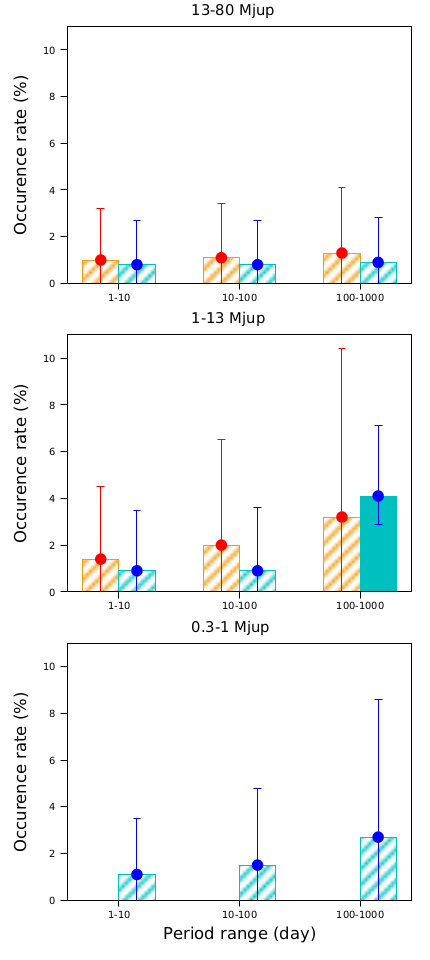}
\caption{Constraints on companion occurrence rates from our combined \sophie~+ \harps~survey. {\it From top to bottom}: brown dwarf companions (13-80 \Mjup), `Jupiter-mass' companions (1-13 \Mjup), `Saturn-mass' companions (0.3-1 \Mjup). {\it From left to right on each row}: period ranges of 1-10, 10-100, and 100-1000 days. {\it Cyan}: stellar mass $\leq$~1.5~\Msun~subsample (with 1$\sigma$ error bars in blue). {\it Orange}: stellar mass $>$~1.5~\Msun~subsample (with 1$\sigma$ error bars in red). Actual occurrence rates are displayed in full color ($n_{\rm det}$ $\geq$ 1) and upper limits ($n_{\rm det}$ = 0) are displayed in hatch.}
\label{occur}
\end{figure}

Our upper limits on the BD occurrence rate around AF stars ($\leq$ 4\%~at 1$\sigma$) are compatible with the results obtained for solar-mass stars. The close BD occurrence rate for solar-mass stars is estimated to be $<$ 1\%~for $P <$~5~yr \citep[or sma $\lesssim$ 3 au, see][]{grether06} and around 0.6\%~for $P \lesssim$ 1~yr \citep{sahlmann11,grieves17}. \cite{jones16} found an occurrence rate of $1.2^{1.5}_{-0.4}$\%~for BD companions within 5 au around evolved intermediate-mass giant stars off the Main-Sequence ($1.6^{2}_{−0.5}$\%~when restricted to stars $\geq$ 1.5 \Msun), which is also compatible with our constrains. Interestingly, these authors did not detect any BD companion within 2 au from their targets, in agreement with our results. In short, we do not find any significant difference so far in terms of close BD companion frequency between AF dwarfs and FGK dwarfs, and between AF dwarfs and GK subgiants.

\item We then consider `Jupiter-mass' companions (1 $\leq$ \msini~$\leq$ 13 \Mjup). The GP system occurrence rate for our full AF sample in the 1-$10^{3}$ day range is $2.5_{-0.7}^{+2}$\%, which is slightly lower than (but still compatible with) the $3_{-0.6}^{+4.7}$\%~value we obtained in Paper IX. If looking at our two stellar mass subsamples separately, we find that the five planetary-mass companions with periods within $10^{3}$ days we considered in this analysis all happen to orbit around stars $<$ 1.5 \Msun. For this low-\Mstar~subsample, our survey completeness is close to 90\%~(almost the same as in Paper IX); the occurrence rate is $3.7_{-1.1}^{+2.8}$\%~($P \leq 10^{3}$ days), which is very close to the value we found in Paper IX, but with a significantly lower uncertainty. This result is compatible with the $\sim 4 \pm 1$\%~GP frequency found by \cite{cumming08} for GP above 1 \Mjup~with sma below 2 au orbiting around FGK solar-like stars. 

\renewcommand{\arraystretch}{1.25}
\begin{table*}[t!]
\caption{GP occurrence rate around AF dwarf stars. The parameters are displayed in normal, bold or italic fonts when considering the full star sample, the most massive ($>$ 1.5 \Msun) stars only or the least massive ($\leq$ 1.5 \Msun) stars only, respectively. For \msini~in the 0.3 to 1-\Mjup~range, we display the GP occurrence rate and other parameters only for our low-\Mstar~subsample, as our completeness is almost null in this domain for our higher-mass targets. The sixth column gives the value derived from the binomial statistics based on $n_{\rm det}$ or $n_{\rm det}^{'}$, while the ninth column gives the corresponding GP occurrence rate or upper limit at a 1$\sigma$ uncertainty.}
\label{tab_occur}
\begin{center}
\begin{tabular}{l c c c c c l l l}\\
\hline
\hline
\msini    & Orbital period  & Search       & Detected    &  Missed       & GP occ. rate    & \multicolumn{2}{c}{Confidence intervals} & GP occ. rate \\
interval  & interval        & completeness & GP systems  & GP systems    & (computed value)& $1\sigma$ & $2\sigma$   &     \\
(\Mjup)   & (day)           &  $C$  (\%)   & $n_{\rm det}$ & $n_{\rm miss}$  & (\%)            & (\%)      &  (\%)       & (\%)      \\
\hline
\hline
 13-80    &  1-10           & 99             &   0    &  $\sim$0      & 0.5     & 0 - 1.5      & 0 - 2.5           & $\leq$ 1.5 \\
 (BD)     &                 & {\bf 98}       & {\bf 0}&  {\bf $\sim$0}&{\bf 1}  &{\bf 0 - 3.2} & {\bf 0 - 5.5}     & {\bf $\leq$ 3.2}\\
          &                 & {\it $\sim$99} & {\it 0}&  {\it $\sim$0}&{\it 0.8}&{\it 0 - 2.7} &{\it 0 - 4.6}      & {\it $\leq$ 2.7}\\
\hline
          &  10-100         & 95             &   0    &  $\sim$0      & 0.5     & 0 - 1.5      & 0 - 2.6           & $\leq$ 1.5 \\
          &                 & {\bf 92}       & {\bf 0}&  {\bf 0.1}    &{\bf 1.1}&{\bf 0 - 3.4} & {\bf 0 - 5.9}     & {\bf $\leq$ 3.4} \\
          &                 & {\it 98}       & {\it 0}&  {\it $\sim$0}&{\it 0.8}&{\it 0 - 2.7} & {\it 0 - 4.7}     & {\it $\leq$ 2.7} \\
\hline
          &  100-1000       & 87             &   0    &  0.2          & 0.5     & 0 - 1.7      & 0 - 2.9           & $\leq$ 1.7 \\
          &                 & {\bf 78}       & {\bf 0}&  {\bf 0.3}    &{\bf 1.3}&{\bf 0 - 4.1} & {\bf 0 - 7}       & {\bf $\leq$ 4.1} \\
          &                 & {\it 95}       & {\it 0}&  {\it 0.1}    &{\it 0.9}&{\it 0 - 2.8} & {\it 0 - 4.8}     & {\it $\leq$ 2.8} \\
\hline
          & 1-100           & 97             &   0    & $\sim$0       & 0.5     & 0 - 1.5      & 0 - 2.6           & $\leq$ 1.5     \\
          &                 & {\bf 95}       & {\bf 0}& {\bf 0.1}     &{\bf 1}  &{\bf 0 - 3.3} & {\bf 0 - 5.7}     & {\bf $\leq$ 3.3} \\
          &                 & {\it 99}       & {\it 0}& {\it $\sim$0} &{\it 0.8}&{\it 0 - 2.7} & {\it 0 - 4.6}     & {\it $\leq$ 2.7} \\
\hline
          &  1-1000         & 94             &   0    & 0.1           & 0.5     & 0 - 1.6      & 0 - 2.7           & $\leq$ 1.6  \\
          &                 & {\bf 89}       & {\bf 0}& {\bf 0.1 }    &{\bf 1.1}&{\bf 0 - 3.6} & {\bf 0 - 6}       & {\bf $\leq$ 3.6} \\
          &                 & {\it 97}       & {\it 0}& {\it $\sim$0} &{\it 0.9}&{\it 0 - 2.8} &{\it 0 - 4.7}      & {\it $\leq$ 2.8} \\
\hline
\hline
 1-13         &  1-10       & 84             &  0     &  0.2          & 0.5      & 0 - 1.8      & 0 - 3            & $\leq$ 1.8 \\
(``Jupiters'')&             & {\bf 71}       & {\bf 0}& {\bf 0.4}     &{\bf 1.4} &{\bf 0 - 4.5} & {\bf 0 - 7.6}    & {\bf $\leq$ 4.5} \\
              &             & {\it 95}       & {\it 0}&  {\it 0.1}    &{\it 0.9} &{\it 0 - 2.8} &{\it 0 - 4.8}     & {\it $\leq$ 2.8} \\
\hline
              &  10-100     & 72             &  0     &  0.4          & 0.6      & 0 - 2.1      & 0 - 3.5          & $\leq$ 2.1 \\
              &             & {\bf 49}       & {\bf 0}& {\bf 1}       &{\bf 2}   &{\bf 0 - 6.5} & {\bf 0 - 11}     & {\bf $\leq$ 6.5} \\
              &             & {\it 91}       & {\it 0}& {\it 0.1}     &{\it 0.9} &{\it 0 - 2.9} & {\it 0 - 5}      & {\it $\leq$ 2.9} \\
\hline
              &  100-1000   & 58             &  4     &  2.9          & 3.1      & 2.2 - 5.5    & 1.2 - 7.9        & $3.1^{+2.4}_{-0.9}$ \\
              &             & {\bf 30}       & {\bf 0}& {\bf 2.3}     & {\bf 3.2}&{\bf 0 - 10.4}&{\bf 0 - 17.7}    & {\bf $\leq$ 10.4} \\
              &             & {\it 82}       & {\it 4}& {\it 0.9}     & {\it 4.1}&{\it 2.9 - 7.1}&{\it 1.6 - 10.2} & {\it $4.1^{+3}_{-1.2}$} \\
\hline
              & 1-100       & 78             &  0     &  0.3          & 0.6      & 0 - 1.9      &  0 - 3.2         & $\leq$ 1.9     \\
              &             & {\bf 60}       & {\bf 0}&  {\bf 0.7}    &{\bf 1.6} &{\bf 0 - 5.3} & {\bf 0 - 9}      & {\bf $\leq$ 5.3} \\
              &             & {\it 93}       & {\it 0}&  {\it 0.1}    &{\it 0.9} &{\it 0 - 2.9} & {\it 0 - 4.9}    & {\it $\leq$ 2.9} \\
\hline
              &  1-1000     & 71             &  4     &  1.6          & 2.5      & 1.8 - 4.5    & 1 - 6.4          & $2.5^{+2}_{-0.7}$    \\
              &             & {\bf 50}       & {\bf 0}&  {\bf 1}      & {\bf 2}  &{\bf 0 - 6.3} &{\bf 0 - 10.8}    & {\bf $\leq$ 6.3} \\
              &             & {\it 89}       & {\it 4}&  {\it 0.5}    & {\it 3.7}&{\it 2.6 - 6.5}&{\it 1.5 - 9.3}  & {\it $3.7^{+2.8}_{-1.1}$}\\
\hline
\hline
 0.3-1 (``Saturns'') &  1-10       & {\it 76} & {\it 0} &  {\it 0.3} & {\it 1.1} &{\it 0 - 3.5}&{\it 0 - 6}       & {\it $\leq$ 3.5}\\
\hline
                     &  10-100     & {\it 56} & {\it 0} &  {\it 0.8} & {\it 1.5} &{\it 0 - 4.8}&{\it 0 - 8.2}     & {\it $\leq$ 4.8} \\
\hline
                     & 100-1000    & {\it 31} & {\it 0} &  {\it 2.2} & {\it 2.7} &{\it 0 - 8.6}&{\it 0 - 14.7}    & {\it $\leq$ 8.6} \\
\hline
                     & 1-100       & {\it 66} & {\it 0} &  {\it 0.5} &{\it 1.3}  &{\it 0 - 4.1}&{\it 0 - 6.9}     & {\it $\leq$ 4.1} \\
\hline
                     &  1-1000     & {\it 54} & {\it 0} &  {\it 0.8} &{\it 1.5}  &{\it 0 - 4.9}&{\it 0 - 8.4}     & {\it $\leq$ 4.9}  \\
\hline
\hline
\end{tabular}
\end{center}
\end{table*}
\renewcommand{\arraystretch}{1.}

For our high-mass subsample, we do not detect any GP and get a completeness of 50\%. We thus constrain the 1-13~\Mjup~GP system occurrence rate around A-type stars to be $\lesssim$ 6\%~(1$\sigma$) for $P \leq 10^{3}$ days. This upper limit is two to three times lower than the upper limit we obtained in Paper IX (in agreement with a sample twice larger). Still, we cannot infer from this result if the GP occurrence rate around A-type stars is significantly different from the value we derive for our F-type targets, or from the GP occurrence rate provided by \cite{cumming08} for FGK dwarf stars. We can only say that the occurrence rate of ``super-Jupiter'' GP does not steeply rise for A-type dwarfs compared to FGK dwarfs for $P \leq 10^{3}$ days. Given the remaining uncertainty, it is still compatible with the increased global GP frequency versus stellar mass predicted by the core-accretion theory \citep[][see also below]{kennedy08}.

\item Finally, we consider ``Saturn-mass'' (0.3 $\leq$ \msini~$\leq$ 1 \Mjup) companions around our lower-mass targets ($\leq$ 1.5 \Msun). We do not detect any such companion in our \sophie~survey. It was already the case for our \harps~survey (Paper IX). With a completeness between 75\%~(1 $\leq P \leq$ 10 days) and 40\%~($10^{2} \leq P \leq 10^{3}$ days), we constrain the occurrence rate to be $\leq$ 3.5\%~and $\leq$ 8.6\%~(1$\sigma$), respectively. These upper limits are roughly twice lower than those we obtained in Paper IX for the \harps~survey only.
\end{enumerate}

\paragraph{Companion frequency vs period --}

For BD-mass companions, our completeness is above 90\%~for all the considered period ranges between 1 and $10^{3}$ days; thus there is no really significant difference between the occurrence rate values we get for these period ranges.

In the case of planetary-mass companions, five out of the six GP we detected (at the exception of the candidate HD~113337c with its longer, $\sim$3260-day, period) belong to a somewhat limited ($P$,~\msini) domain, with orbital periods between 200 and 1000 days and minimal masses between 1 and 10 \Mjup. Our sensitivity to companions in this domain ranges from 20-30\%~to 70-80\%; we achieve a better completeness closer to the star, yet we do not detect any GP for these smaller periods. We remind that \safir~preliminary tests based on \elodie~RV had shown to be able to re-detect a known $P$$\sim$3-day Hot Jupiter \citep{galland05b} and to detect a $P$$\sim$28-day BD \citep[][]{galland06}. Thus, this apparent `concentration' of GP at the largest considered period ranges is most probably not induced by a bias in our method.\\

In terms of companion frequency, we constrain the Hot Jupiter ($P$ $<$ 10 or $<$ $10^{2}$ days) occurrence rate to be below $\leq$ 3\%~for our targets $\leq$ 1.5 \Msun~and to be below $\leq$ 4.5-6.5\%~for our targets $>$ 1.5 \Msun. These upper limits on the occurrence rates and their 1$\sigma$ uncertainties are decreased by a factor $\sim$2 compared to Paper IX. In the case of our low-mass targets, this ``Hot Jupiter'' upper limit is smaller than (but still compatible with) the $4.1_{-1.2}^{+3}$\%~value we get in the $10^{2}$-$10^{3}$-day range (where most of our detected GP are located). This result is compatible with a possible increase of the GP frequency with the orbital period, at least for our low-mass subsample. Based on the core-accretion theory, \cite{kennedy09} predicted that GP would be found at longer periods around more massive stars. The fact that we do not detect any GP around our high-mass targets in the 1-$10^{3}$-day range could be in agreement with such a trend (although it remains speculative as long as we are not able to constraint the occurrence rate of GP on wider ($\sim$3-5 au) orbits around A-type stars (with the help of direct imaging surveys and/or Gaia).\\
 
Since we do not detect any Hot Jupiter in our sample, we only have an upper limit on the Hot Jupiter frequency, \ie~$\leq$~3-4.5\%~(1$\sigma$). Then, we cannot tell if there is a difference with the Hot Jupiter rates of 0.5 to 1.5\%~derived for FGK MS stars \citep[][]{santerne16}. Since several Hot Jupiters have recently been detected by transits and Doppler tomography around AF stars \citep[see \eg][]{bourrier15,stevens16,crouzet17,mcleod17,zhou17}, it would be highly interesting to compare our constraints on the Hot Jupiter frequencies around AF stars with similar statistics derived from transit survey results.\\

\paragraph{Comparison to other stars --}

The GP occurrence rate ($3.7_{-1.1}^{+2.8}$\%) we derive for our F-type (1 to 1.5 \Msun) subsample shows no significant difference when compared to solar-like FGK stars. However, our upper limit on the GP occurrence rate ($\leq$ 6.3\%~at 1$\sigma$) around A-type stars (1.5-2.5 \Msun) is significantly smaller than both the GP frequencies around GK subgiants ($11 \pm 2$\%) and giants ($\simeq 15_{-3}^{+8}$\%) derived in approximately the same period ranges by \cite{johnson10} and \cite{reffert14}, respectively. Since we compute our detection limits only for circular orbits (as it is the common approach; see Paper IX for more details), the uncertainties on our companion occurrence rates are slightly underestimated by not considering a variable companion eccentricity, and might still be compatible with the above results. Still, this discrepancy in the GP occurrence rate between our A-type subsample and the evolved star samples of \cite{johnson10} and \cite{reffert14} remains significant. A possible interpretation is that different stellar properties (stellar mass, metallicity, ...) significantly impact the formation and evolution of giant planets \citep[see \eg][]{ghezzi18}.

Finally, it is also interesting to compare the companion frequency around AF MS stars to the companion population around white dwarfs (WD), as the latter typically have AF-type progenitors. Based on searches for an infrared excess in the WD spectral energy distribution, \cite{farihi05} and \cite{girven11} estimated the global BD occurrence rate around WD to be $\lesssim 1$\%~(this estimation should be robust, as WD are ideal targets to search for low-mass stellar and substellar companions, due to their low luminosity). This result is consistent with the overall 3-4\%~upper limit we obtain on the BD occurrence rate in the 1-$10^{3}$-day period range. It is also in agreement with the absence of BD detections within 2 au from the primary reported by \cite{jones16} for subgiant stars. We note that our upper limit for the considered periods ($P = 10^{3}$ days, or sma $\simeq$ 2.5-3 au) roughly corresponds to the closest orbits for which a substellar companion is expected to avoid engulfment during the red giant transition \citep{mustill12,nordhaus13}.

\section{Conclusion}\label{sect:conclu}

We have carried out a systematic RV search for BD and GP companions around 125 northern AF MS dwarfs with \sophie. This survey has led to the detection of one (possibly two) GP around the F6V dwarf HD~113337 and does not contradict the existence of the already reported GP around HD~16232. We additionally detected fourteen high-mass and/or distant companions based on high-amplitude RV variations or long-term RV trends, of which twelve are probably of stellar nature, based either on CCF and/or FWHM variations or independent detection from the literature. We finally detected one (possibly two) faint SB2 binary based on a strong RV-FWHM correlation and complex low-amplitude RV variations (HD~191195 and hypothetically HD~185395).\\

We combined this \sophie~survey to the \harps~twin survey detailed in Paper IX to conduct a global statistical study of the BD/GP occurrence rate within $\sim$2-2.5 au around AF MS stars. We were able to derive good constraints on these occurrence rates, with a BD occurrence rate below 3-4\%~and a GP (1~$<$ \msini~$<$ 13 \Mjup) occurrence rate below 3 to 5\%~for periods below 100 days and of $\simeq 4_{-1}^{+3}$\%~for periods in the $10^{2}$-$10^{3}$ day range and stellar masses below 1.5 \Msun. Our BD occurrence rates are compatible with the BD frequencies reported both for FGK dwarfs and for evolved GK subgiants and giants. Our GP occurrence rates for stars with masses in the 1-1.5 \Msun~range are compatible with the GP frequencies reported for FGK dwarfs. However, our upper limits on the GP frequency around stars with masses in the 1.5-2.5 \Msun~range are somewhat smaller than the GP frequencies reported for evolved GK stars, which are considered to be their descendants.

The fact that we do not detect any GP around our A-type targets (compared to the six GP detected or reported for our F-type targets) is compatible with the prediction that GP form further away from the primary and migrate less with an increasing stellar mass, though it does not confirm it. Our results can be of interest for planetary population synthesis studies \citep[see \eg][]{mordasini09,mordasini12}, especially when considering the influence of the stellar mass on the planetary formation and migration processes. In this context, the astrometry from Gaia will probe with an unprecedented sensitivity the GP population at a few au around A-type stars in the solar neighborhood, thus allowing a promising synergy with our RV survey \citep{sozzetti15,sahlmann16}.

Finally, we did not detect any Hot Jupiter out of our 225-target sample, constraining their occurrence rate to be below 3 to 5\%~if considering A and F stars separately. While this result does not show a significant difference with the results reported for solar-mass stars, it would be of great interest to compare it to a statistical study of the Hot Jupiters that have recently been detected by transits around more distant AF stars.\\

\begin{acknowledgements}
We acknowledge support from the French CNRS and from the Agence Nationale de la Recherche (ANR grant GIPSE ANR-14-CE33-0018). NCS acknowledges support by Funda\c{c}\~ao para a Ci\^encia e a Tecnologia (FCT, Portugal) through national funds and by FEDER through COMPETE2020 by grants UID/FIS/04434/2013 \& POCI-01-0145-FEDER-007672 and PTDC/FIS-AST/1526/2014 \& POCI-01-0145-FEDER-016886, as well as through Investigador FCT contract nr. IF/00169/2012/CP0150/CT0002. These results have made use of the SIMBAD database, operated at the CDS, Strasbourg, France. We would like to thank B. Gaensick and J. Farihi for their useful suggestions on the companion occurrence rate of white dwarfs as the descendents of AF MS stars. We would like to thank our anonymous referee for the useful additional suggestions.
\end{acknowledgements}
\bibliographystyle{aa}
\bibliography{SAF_planets}

\Online
\begin{appendix}

\section{\sophie~sample}\label{app:sample}

The main physical properties and the main characteristics of the spectroscopic observables for each of the targets of our \sophie~sample are detailed in Table.

\begin{center}
\onecolumn
\setlength\tabcolsep{3.5pt}
\begin{longtable}[1]{l l c c c c | l p{0.5cm} c c c c c c l l l l}\\
\caption{\label{tab:sample} Stellar characteristics and detailed results for the 125 targets of our \sophie~RV survey. Spectral type (ST) and \bv~values are taken from the CDS database, as well as the \vsini~values. Stellar masses are taken from \cite{allende99}. The survey results include the observation time baseline (TBL), the number of computed spectra $N_{\rm m}$, the peak-to-peak (i.e., maximum-to-minimum) amplitude {\it A}, rms and mean uncertainty $<${\it U}$>$ on the RV and BIS measurements, the RV-BIS correlation (Pearson’s coefficient), and the mean FWHM ($<$FW$>$). V stands for our RV variability criterion (Paper IX), with V for RV variable stars and C for RV constant targets. The CL column reports the correction applied on the RV, if any ($\dagger$: correction from binary or planetary fit; $\star$: correction from RV-BIS activity correlation).}\\
\hline
\multicolumn{6}{c|}{Stellar characteristics} & \multicolumn{12}{c}{Survey detailed results.}\\
\hline
HD & HIP & ST & \bv & \vsini  & Mass          & TBL & $N_{\rm m}$   & \multicolumn{3}{c}{RV}        & \multicolumn{3}{c}{BIS}      & RV-    & $<$FW$>$  & V & CL \\

   &     &    &     &         &               &     &             & \multicolumn{3}{c}{\raisebox{.5\baselineskip}{$\overbrace{\hspace{2.7cm}}$}} & \multicolumn{3}{c}{\raisebox{.5\baselineskip}{$\overbrace{\hspace{2.7cm}}$}} & BIS & & & \\

   &     &    &     &         &               &     &             & {\it A} &rms  & $<${\it U}$>$ & {\it A} & rms &$<${\it U}$>$ & corr.    &          &    &    \\

   &     &    &     &         &               &     &             & \multicolumn{3}{c}{\raisebox{.5\baselineskip}{$\underbrace{\hspace{2.7cm}}$}} & \multicolumn{3}{c}{\raisebox{.5\baselineskip}{$\underbrace{\hspace{2.7cm}}$}} & & & & \\

   &     &    &     & \kms    & \Msun         & day &            & \multicolumn{3}{c}{\ms}        & \multicolumn{3}{c}{\ms}      &   &  \kms       &   &    \\
\hline
\hline
\endfirsthead
\caption{Continued.}\\
\hline
\multicolumn{6}{c|}{Stellar characteristics} & \multicolumn{12}{c}{Survey detailed results.}\\
\hline
HD & HIP & ST & \bv & \vsini  & Mass          & TBL & $N_{\rm m}$   & \multicolumn{3}{c}{RV}        & \multicolumn{3}{c}{BIS}      & RV-    & $<$FW$>$  &  V & CL \\

   &     &    &     &         &               &     &             & \multicolumn{3}{c}{\raisebox{.5\baselineskip}{$\overbrace{\hspace{2.7cm}}$}} & \multicolumn{3}{c}{\raisebox{.5\baselineskip}{$\overbrace{\hspace{2.7cm}}$}} & BIS &  & & \\

   &     &    &     &         &               &     &                & {\it A} &rms  & $<${\it U}$>$ & {\it A} & rms &$<${\it U}$>$ & corr.    &             &     &    \\

   &     &    &     &         &               &     &             & \multicolumn{3}{c}{\raisebox{.5\baselineskip}{$\underbrace{\hspace{2.7cm}}$}} & \multicolumn{3}{c}{\raisebox{.5\baselineskip}{$\underbrace{\hspace{2.7cm}}$}} & & &  & \\

   &     &    &     & \kms    & \Msun         & day  &                & \multicolumn{3}{c}{\ms}        & \multicolumn{3}{c}{\ms}     &   &  \kms     &    &    \\
\hline
\hline
\endhead
\hline
\endfoot
400    & 699    & F8IV  & 0.456 & 3   & 1.15 & 1503 & 34 & 40.6 & 9.8  & 5.6  & 29.9 & 7.7  & 6.3  & -0.37 & 11.1 & C &  \\
1404   & 1473   & A2V   & 0.05  & 110 & 2.21 & 1134 & 16 & 1280 & 338  & 259  &      &      &      &       &      & C &  \\
3268   & 2832   & F7V   & 0.52  & 5.8 & 1.37 & 1493 & 21 & 36.1 & 10.7 & 6.4  & 22.9 & 5.4  & 9.9  & 0.02  & 10.7 & C &  \\
3440   & 3132   & F6V   & 0.501 & 4.7 & 1.19 & 173  & 17 & 32.2 & 7.9  & 5.5  & 23.9 & 5.2  & 5.5  & -0.12 & 9.4  & C &  \\
6288   & 4979   & A8IV  & 0.243 & 94  & 1.76 & 361  & 16 & 743  & 219  & 82.1 & 6114 & 1350 & 205  & -0.17 & 134  & V &  \\
6961   & 5542   & A7V   & 0.18  & 91  & 1.8  & 1417 & 14 & 198  & 46.1 & 48.4 &      &      &      &       &      & C &  \\
7193   & 5631   & F5V   & 0.49  & 8.2 & 1.05 & 357  & 16 & 67.3 & 22.3 & 6.6  & 63.7 & 20.6 & 10.8 & -0.66 & 14.3 & V &  \\
8723   & 6706   & F2V   & 0.343 & 60  & 1.38 & 1553 & 25 & 263  & 65.7 & 38.8 & 5293 & 1167 & 96.1 & -0.37 & 88.6 & C &  \\
8829   & 6748   & F0.5V & 0.283 & 18  & 1.49 & 102  & 16 & 251  & 89.6 & 12.5 & 123  & 28.9 & 28.6 & -0.4  & 28.8 & V &  \\
8907   & 6878   & F8V   & 0.49  & 9.5 & 1.23 & 395  & 19 & 219  & 55.5 & 8.3  & 118  & 34.1 & 16.3 & -0.56 & 20.7 & V &  \\
9780   & 7447   & F0IV  & 0.24  & 113 & 1.8  & 823  & 13 & 888  & 265  & 102  & 6261 & 2129 & 255  & 0.58  & 177  & V &  \\
10453  & 7916   & F5V   & 0.44  & 13  & 1.49 & 1589 & 17 & 2244 & 660  & 10.9 & 19133& 4783 & 23.9 & 0.92  & 22.8 & V & $\dagger$ \\
       &        &       &       &     &      &      &    & 71.9 & 21.6 &      &      &      &      &       &      & C &  \\
11973  & 9153   & F0V   & 0.271 & 97  & 1.77 & 648  & 21 & 926  & 236  & 59.9 & 3876 & 1173 & 149  & -0.2  & 138  & V &  \\
12111  & 9480   & A3V   & 0.16  & 69  & 1.97 & 1399 & 14 & 202  & 48.8 & 35.5 & 3015 & 927  & 87.8 & 0.32  & 86.5 & C &  \\
13555  & 10306  & F5V   & 0.4   & 9   & 1.45 & 679  & 39 & 80.7 & 21.1 & 6.1  & 104  & 25   & 8.4  & 0.05  & 13.5 & V &  \\
16232  & 12184  & F6V   & 0.5   & 42  & 1.2  & 1892 & 27 & 568  & 161  & 21   & 700  & 160  & 50.9 & -0.08 & 60.7 & V & $\dagger$ \\
       &        &       &       &     &      &      &    & 262  & 72.5 &      &      &      &      &       &      & V &  \\
16765  & 12530  & F7V   & 0.52  & 32  & 1.2  & 318  & 30 & 276  & 60.1 & 27.6 & 556  & 129  & 67.8 & -0.51 & 48.2 & V &  \\
16895  & 12777  & F8V   & 0.51  & 9   & 1.15 & 687  & 23 & 62.4 & 16.5 & 7.2  & 39   & 10   & 12.8 & 0.33  & 14.9 & V &  \\
17948  & 13665  & F5V   & 0.381 & 6.1 & 1.33 & 2102 & 24 & 62.1 & 12.5 & 6.6  & 45.9 & 11.4 & 10.5 & -0.3  & 13.5 & C &  \\
18404  & 13834  & F5IV  & 0.41  & 26  & 1.35 & 1360 & 14 & 110  & 28.8 & 11.9 & 586  & 198  & 27   & -0.73 & 37.7 & V &  \\
20395  & 15244  & F5V   & 0.364 & 6   & 1.36 & 1175 & 22 & 6444 & 2669 & 6.6  & 587  & 238  & 10.6 & -0.97 & 13.2 & V & $\dagger$ \\
       &        &       &       &     &      &      &    & 46.3 & 14.5 &      &      &      &      &       &      & V &  \\
20677  & 15648  & A3V   & 0.07  & 130 & 2    & 8    & 22 & 1445 & 299  & 178  &      &      &      &       &      & C &  \\
25490  & 18907  & A0.5V & 0.045 & 65  & 2.27 & 1817 & 14 & 741  & 197  & 131  &      &      &      &       &      & C &  \\
25621  & 18993  & F6V   & 0.5   & 15  & 1.45 & 1433 & 15 & 65.5 & 21.1 & 7.8  & 137  & 32.8 & 14.9 & -0.73 & 25.5 & V & $\star$ \\
       &        &       &       &     &      &      &    & 51.8 & 14.5 &      &      &      &      & 0     &      & C &  \\
27819  & 20542  & A2V   & 0.15  & 45  & 1.96 & 1399 & 25 & 633  & 182  & 17.5 & 943  & 247  & 41.9 & -0.05 & 75.3 & V &  \\
27934  & 20635  & A7IV-V& 0.143 & 87  & 2.07 & 1883 & 18 & 1323 & 327  & 43.2 &      &      &      &       &      & V &  \\
27946  & 20641  & A7V   & 0.231 & 175 & 1.82 & 1750 & 17 & 1104 & 272  & 254  &      &      &      &       &      & C &  \\
27962  & 20648  & A2IV-V& 0.049 & 8.3 & 2.29 & 1009 & 70 & 78.6 & 19   & 6.4  & 91.7 & 14.1 & 10   & 0.09  & 18   & V &  \\
28355  & 20901  & A5V   & 0.212 & 93  & 1.91 & 1535 & 20 & 408  & 89   & 42.7 & 36564& 6306 & 106  & 0.36  & 115  & V &  \\
29488  & 21683  & A5V   & 0.157 & 115 & 2.04 & 1370 & 26 & 1085 & 271  & 79.2 &      &      &      &       &      & V &  \\
31662  & 23380  & F4V   & 0.391 & 28  & 1.44 & 2160 & 33 & 120  & 29.5 & 13   & 647  & 167  & 29.9 & -0.22 & 42.5 & V &  \\
31675  & 23484  & F6V   & 0.48  & 8.8 & 1.23 & 645  & 32 & 110  & 31.4 & 6.6  & 150  & 32.7 & 10.6 & -0.51 & 19.2 & V &  \\
33608  & 24162  & F5V   & 0.423 & 14  & 1.43 & 2216 & 14 & 116  & 32.4 & 8    & 195  & 67   & 15.5 & -0.84 & 22.9 & V & $\star$ \\
       &        &       &       &     &      &      &    & 68.2 & 17.4 &      &      &      &      & 0     &      & V &  \\
43042  & 29650  & F5IV-V& 0.44  & 10  & 1.33 & 1018 & 49 & 65.8 & 13.6 & 7.2  & 40   & 7.5  & 12.9 & -0.01 & 11.2 & C &  \\
43318  & 29716  & F5V   & 0.5   & 4.6 & 1.4  & 1850 & 21 & 33.7 & 10.2 & 5.8  & 32.8 & 8.1  & 7.2  & 0.42  & 10.6 & C & $\dagger$ \\
       &        &       &       &     &      &      &    & 20.8 & 5.2  &      &      &      &      &       &      & C &  \\
43386  & 29800  & F5V   & 0.42  & 18  & 1.33 & 2247 & 21 & 90.3 & 23.8 & 9.8  & 258  & 66.3 & 21   & -0.55 & 29.9 & V &  \\
48737  & 32362  & F5IV-V& 0.43  & 68  & 1.63 & 1448 & 13 & 123  & 37.1 & 28.5 & 1163 & 393  & 70.1 & -0.15 & 92   & C &  \\
58461  & 35998  & F5V   & 0.42  & 14  & 1.49 & 1861 & 22 & 371  & 114  & 10.1 & 837  & 252  & 21.6 & -0.93 & 24.4 & V & $\star$ \\
       &        &       &       &     &      &      &    & 186  & 42.3 &      &      &      &      & 0     &      & V &  \\
58855  & 36439  & F6V   & 0.414 & 10  & 1.25 & 793  & 69 & 49.7 & 10.6 & 5.9  & 55.1 & 10   & 7.8  & -0.02 & 14.9 & C &  \\
58946  & 36366  & F1V   & 0.32  & 63  & 1.5  & 456  & 18 & 372  & 86.7 & 35.8 &      &      &      &       &      & V &  \\
63332  & 38325  & F6V   & 0.441 & 7   & 1.13 & 1136 & 24 & 63.2 & 15   & 6.1  & 52.9 & 13.4 & 8.7  & 0.31  & 14.7 & V &  \\
69548  & 40875  & F4V   & 0.367 & 55  & 1.34 & 1500 & 24 & 298  & 62.9 & 29.1 & 1923 & 462  & 71.5 & -0.33 & 76.2 & V &  \\
69897  & 40843  & F6V   & 0.51  & 5   & 1.25 & 1193 & 27 & 56.9 & 13.5 & 5.7  & 38.8 & 9.2  & 6.8  & 0.03  & 10.9 & V &  \\
70313  & 41152  & A3V   & 0.12  & 100 & 1.83 & 1004 & 13 & 650  & 221  & 143  & 77613& 20230& 358  & -0.15 & 169  & C &  \\
75332  & 43410  & F7V   & 0.498 & 10  & 1.17 & 3412 & 68 & 125  & 33.4 & 6.1  & 95.5 & 20.6 & 8.6  & -0.45 & 15.2 & V &  \\
75616  & 43625  & F5V   & 0.42  & 7.5 & 1.04 & 826  & 37 & 149  & 35.3 & 7    & 114  & 31.5 & 12.3 & -0.49 & 16.5 & V &  \\
76398  & 43932  & A7IV  & 0.161 & 119 & 1.98 & 2246 & 28 & 503  & 143  & 85   & 77355& 14173& 212  & -0.1  & 147  & C &  \\
76582  & 44001  & F0IV  & 0.207 & 92  & 1.71 & 754  & 16 & 457  & 123  & 64.9 & 9508 & 2930 & 162  & -0.08 & 130  & C &  \\
76644  & 44127  & A7V   & 0.19  & 129 & 1.64 & 1163 & 20 & 1470 & 443  & 84   & 35467& 8604 & 210  & 0.19  & 188  & V &  \\
78154  & 45038  & F7V   & 0.49  & 5.4 & 1.43 & 765  & 82 & 49.1 & 10.4 & 5.4  & 33   & 6.2  & 5.2  & -0.01 & 12   & C &  \\
79439  & 45493  & A6V   & 0.175 & 145 & 1.89 & 742  & 34 & 1348 & 331  & 135  &      &      &      &       &      & V &  \\
80290  & 45836  & F3V   & 0.42  & 4.5 & 1.14 & 2670 & 40 & 40.2 & 9.6  & 5.8  & 32.2 & 7.4  & 7.4  & -0.25 & 10.6 & C &  \\
82328  & 46853  & F7V   & 0.46  & 6.3 & 1.53 & 3367 & 75 & 102  & 17   & 5.7  & 106  & 13.3 & 14.3 & 0.04  & 13.9 & V & $\dagger$ \\
       &        &       &       &     &      &      &    & 66.5 & 11.7 &      &      &      &      &       &      & V &  \\
89449  & 50564  & F6IV-V& 0.44  & 16  & 1.44 & 3375 & 47 & 112  & 25.1 & 9.1  & 372  & 75.1 & 22.6 & -0.43 & 25.4 & V &  \\
90277  & 51056  & F0V   & 0.242 & 33  & 2.1  & 1171 & 30 & 169  & 38.4 & 12.7 & 800  & 168  & 29.1 & -0.24 & 50.8 & V &  \\
90470  & 51200  & A3V   & 0.138 & 105 & 1.89 & 2157 & 24 & 2215 & 463  & 222  & 24058& 5268 & 556  & 0.06  & 187  & V &  \\
91312  & 51658  & A7IV  & 0.197 & 128 & 1.83 & 2634 & 34 & 5202 & 1724 & 91.5 &      &      &      &       &      & V & $\dagger$ \\
       &        &       &       &     &      &      &    & 1809 & 417  &      &      &      &      &       &      & V &  \\
95418  & 53910  & A1IV  & -0.02 & 35  & 2.28 & 2619 & 28 & 187  & 47.3 & 28   &      &      &      &       &      & C &  \\
97855  & 55044  & F6V   & 0.413 & 5.7 & 1.32 & 2619 & 23 & 41.9 & 11.8 & 6.5  & 34.9 & 9.6  & 10.4 & -0.42 & 12.8 & C &  \\
99747  & 56035  & F5V   & 0.39  & 8.4 & 1.41 & 1255 & 36 & 49.2 & 11   & 10.1 & 102  & 20.9 & 21.6 & 0.14  & 18.1 & C &  \\
102647 & 57632  & A3V   & 0.09  & 115 & 1.9  & 1145 & 102& 700  & 134  & 111  &      &      &      &       &      & C &  \\
102870 & 57757  & F9V   & 0.55  & 4.4 & 1.36 & 3370 & 87 & 54   & 11.2 & 5.4  & 27   & 5.8  & 13.4 & 0.08  & 10.2 & V &  \\
106591 & 59774  & A2V   & 0.09  & 251 & 2.1  & 1990 & 21 & 1962 & 529  & 498  &      &      &      &       &      & C &  \\
107113 & 59879  & F4V   & 0.43  & 5.8 & 1.32 & 1154 & 36 & 71.8 & 16.1 & 6.6  & 52.3 & 14   & 10.5 & -0.06 & 12.7 & V &  \\
109141 & 61212  & F2V   & 0.338 & 142 & 1.46 & 1392 & 16 & 845  & 279  & 152  & 8004 & 2462 & 380  & 0.87  & 195  & C &  \\
111270 & 62402  & A9V   & 0.2   & 93  & 1.89 & 3309 & 24 & 1031 & 223  & 84.9 &      &      &      &       &      & V &  \\
113337 & 63584  & F6V   & 0.372 & 6   & 1.41 & 3368 & 301& 328  & 68.1 & 7.9  & 148  & 24.4 & 19.8 & -0.32 & 13.7 & V & $\dagger$ \\
       &        &       &       &     &      &      &    & 132  & 22.1 &      &      &      &      &       &      & V &  \\
115810 & 64979  & A7IV  & 0.24  & 98  & 1.78 & 1433 & 27 & 739  & 169  & 83.9 & 8411 & 1766 & 209  & -0.14 & 145  & V &  \\
118232 & 66234  & A4V   & 0.12  & 145 & 2.06 & 1435 & 19 & 1285 & 330  & 116  &115938& 25992& 290  & -0.17 & 221  & V &  \\
119992 & 67103  & F7IV-V& 0.47  & 6.2 & 1.25 & 2621 & 53 & 49   & 12.2 & 11.5 & 61   & 11.6 & 28.8 & -0.14 & 13.8 & C &  \\
121164 & 67782  & A8IV  & 0.187 & 65  & 1.9  & 3371 & 60 & 492  & 121  & 37.3 &      &      &      &       &      & V &  \\
121560 & 68030  & F6V   & 0.56  & 5   & 1    & 2680 & 27 & 40.9 & 10.6 & 7.6  & 27.4 & 6.2  & 19   & 0.11  & 9.9  & C &  \\
124675 & 69483  & A7IV  & 0.23  & 111 & 1.93 & 107  & 17 & 2448 & 810  & 77.5 & 36187& 10492& 193  & -0.35 & 174  & V &  \\
125040 & 69751  & F8V   & 0.49  & 36  & 1.16 & 2677 & 17 & 171  & 45.1 & 21.6 & 437  & 133  & 54.1 & -0.11 & 52.9 & V &  \\
125451 & 69989  & F5IV  & 0.364 & 40  & 1.4  & 1476 & 18 & 168  & 43.5 & 21.9 & 647  & 177  & 53.3 & -0.63 & 58.5 & C &  \\
126141 & 70310  & F5V   & 0.358 & 7   & 1.36 & 2680 & 14 & 112  & 41.7 & 8.6  & 139  & 38.2 & 21.4 & -0.67 & 15.1 & V &  \\
129153 & 71759  & F0V   & 0.216 & 105 & 1.68 & 48   & 22 & 1805 & 439  & 109  & 16573& 3043 & 272  & -0.17 & 159  & V &  \\
130044 & 72066  & F0V   & 0.261 & 15  & 1.55 & 519  & 36 & 245  & 63.9 & 9.5  & 338  & 99.8 & 20.1 & -0.7  & 24.1 & V & $\star$ \\
       &        &       &       &     &      &      &    & 178  & 45.5 &      &      &      &      & 0     &      & V &  \\
132254 & 73100  & F8V   & 0.53  & 5   & 1.29 & 1462 & 19 & 34   & 9.3  & 6    & 29.8 & 9    & 8.2  & -0.3  & 14   & C &  \\
132375 & 73309  & F8V   & 0.5   & 6.2 & 1.38 & 1520 & 14 & 25.3 & 9.3  & 6    & 37.3 & 9.4  & 8.1  & -0.44 & 13.7 & C &  \\
134083 & 73996  & F5V   & 0.43  & 45  & 1.33 & 1505 & 20 & 110  & 28.4 & 19.8 & 372  & 85.6 & 47.7 & -0.23 & 64   & C &  \\
136729 & 75043  & A4V   & 0.115 & 145 & 2.05 & 2956 & 32 & 1517 & 373  & 144  &      &      &      &       &      & V &  \\
137391 & 75411  & F0IV  & 0.291 & 89  & 1.8  & 668  & 21 & 1154 & 322  & 44.6 & 16935& 3174 & 111  & 0.35  & 124  & V &  \\
139389 & 76456  & F5V   & 0.4   & 21  & 1.29 & 1224 & 16 & 287  & 68.3 & 14.7 & 177  & 49.7 & 34.6 & 0.57  & 34.1 & V &  \\
142860 & 78072  & F6IV  & 0.5   & 10  & 1.28 & 2071 & 23 & 49.8 & 14.2 & 6.2  & 69.4 & 15.5 & 8.8  & -0.02 & 17.1 & V &  \\
143584 & 78286  & F0IV  & 0.258 & 68  & 1.58 & 2595 & 18 & 2137 & 575  & 75.8 & 2123 & 634  & 190  & -0.21 & 103  & V & $\dagger$ \\
       &        &       &       &     &      &      &    & 1180 & 291  &      &      &      &      &       &      & V &  \\
147365 & 80008  & F4V   & 0.4   & 77  & 1.38 & 2252 & 14 & 357  & 102  & 37.5 & 7755 & 1685 & 93   & -0.05 & 104  & V &  \\
148048 & 79822  & F5V   & 0.37  & 80  & 1.52 & 2987 & 46 & 1678 & 571  & 56.5 & 9907 & 1582 & 141  & -0.16 & 125  & V & $\dagger$ \\
       &        &       &       &     &      &      &    & 272  & 71.2 &      &      &      &      &       &      & C &  \\
149681 & 80480  & F0V   & 0.24  & 118 & 1.62 & 2891 & 27 & 1942 & 417  & 119  & 6499 & 1707 & 297  & -0.29 & 176  & V &  \\
152303 & 81854  & F4V   & 0.41  & 22  & 1.32 & 1803 & 24 & 278  & 73.7 & 13.2 & 531  & 138  & 30.4 & -0.6  & 36.1 & V &  \\
154431 & 83494  & A6V   & 0.19  & 110 & 1.67 & 3195 & 23 & 561  & 150  & 109  &      &      &      &       &      & C &  \\
156295 & 84183  & A7V   & 0.192 & 103 & 1.66 & 3157 & 20 & 1184 & 331  & 80.3 &      &      &      &       &      & V &  \\
159332 & 85912  & F4V   & 0.45  & 7.2 & 1.41 & 3285 & 29 & 48.9 & 10.9 & 8.3  & 54.9 & 15.1 & 20.7 & -0.26 & 15.7 & C &  \\
162003 & 86614  & F5IV-V& 0.44  & 9   & 1.48 & 2890 & 22 & 1806 & 524  & 8.6  & 492  & 124  & 21.5 & -0.28 & 19.2 & V & $\dagger$ \\
       &        &       &       &     &      &      &    & 126  & 34.4 &      &      &      &      &       &      & V &  \\
168151 & 89348  & F5V   & 0.38  & 10  & 1.45 & 3197 & 23 & 79.5 & 19.5 & 11.5 & 89.3 & 18.1 & 28.7 & -0.46 & 14.4 & C &  \\
184960 & 96258  & F7V   & 0.51  & 6.5 & 1.28 & 1900 & 29 & 66.4 & 14.8 & 6.1  & 44.9 & 9.9  & 8.7  & -0.45 & 14.1 & V &  \\
185395 & 96441  & F3V   & 0.38  & 10  & 1.37 & 3482 & 326& 405  & 82.3 & 7.5  & 82.4 & 12.5 & 18.6 & 0.02  & 11.1 & V & $\dagger$ \\
       &        &       &       &     &      &      &    & 288  & 65.2 &      &      &      &      &       &      & V &  \\
186689 & 97229  & A3IV  & 0.18  & 33  & 1.71 & 3193 & 71 & 513  & 101  & 17.9 & 1738 & 358  & 44.7 & -0.81 & 50   & V & $\star$ \\
       &        &       &       &     &      &      &    & 270  & 59.4 &      &      &      &      & 0     &      & V &  \\
187013 & 97295  & F5IV-V& 0.47  & 7.3 & 1.21 & 3234 & 27 & 92.1 & 17.7 & 8    & 81.8 & 18.9 & 20   & 0.15  & 16   & V &  \\
191195 & 99026  & F5V   & 0.39  & 5.5 & 1.49 & 3191 & 265& 272  & 56.4 & 7.6  & 98   & 17.6 & 18.9 & 0.2   & 12.4 & V &  \\
192985 & 99889  & F5V   & 0.391 & 8.7 & 1.44 & 3236 & 19 & 94   & 22.3 & 10.6 & 167  & 41.6 & 26.6 & -0.5  & 18.6 & V &  \\
193369 & 100108 & A2V   & 0.06  & 90  & 1.99 & 2795 & 37 & 1167 & 280  & 185  &      &      &      &       &      & C &  \\
196524 & 101769 & F5IV  & 0.44  & 48  & 1.92 & 1178 & 19 & 2075 & 601  & 18.3 & 934  & 269  & 43.8 & -0.52 & 62.8 & V & $\dagger$ \\
       &        &       &       &     &      &      &    & 197  & 45   &      &      &      &      &       &      & V &  \\
197373 & 102011 & F6IV  & 0.399 & 27  & 1.34 & 387  & 23 & 262  & 61.5 & 12.9 & 303  & 91.9 & 29.7 & -0.56 & 42.9 & V &  \\
197950 & 102253 & A8V   & 0.21  & 160 & 1.67 & 783  & 21 & 1598 & 428  & 399  &      &      &      &       &      & C &  \\
198390 & 102805 & F5V   & 0.395 & 6.5 & 1.21 & 773  & 25 & 35.5 & 9.6  & 7.3  & 70.1 & 15.5 & 13.3 & 0.28  & 14.2 & C &  \\
199254 & 103298 & A5V   & 0.11  & 145 & 2.02 & 1382 & 15 & 1371 & 340  & 151  &      &      &      &       &      & V & $\dagger$ \\
       &        &       &       &     &      &      &    & 597  & 158  &      &      &      &      &       &      & C &  \\
204153 & 105769 & F0V   & 0.32  & 115 & 1.49 & 340  & 15 & 814  & 255  & 88.3 & 3254 & 748  & 220  & 0.34  & 164  & V &  \\
204414 & 105966 & A1V   & 0.046 & 70  & 2    & 748  & 19 & 459  & 113  & 72.3 & 26815& 5247 & 180  & 0.28  & 124  & C &  \\
206677 & 107302 & A7IV-V& 0.204 & 113 & 1.67 & 1414 & 20 & 1412 & 338  & 92   & 8022 & 1992 & 230  & -0.5  & 168  & V &  \\
209369 & 108535 & F5V   & 0.409 & 24  & 1.62 & 1056 & 24 & 369  & 93.8 & 11.7 & 728  & 204  & 26.2 & -0.75 & 39.9 & V & $\star$ \\
       &        &       &       &     &      &      &    & 229  & 62.5 &      &      &      &      & 0     &      & V &  \\
210715 & 109521 & A5V   & 0.142 & 130 & 1.95 & 1879 & 45 & 2152 & 609  & 111  &105313& 17072& 277  & -0.05 & 185  & V & $\dagger$ \\
       &        &       &       &     &      &      &    & 1363 & 376  &      &      &      &      &       &      & V &  \\
211976 & 110341 & F5V   & 0.421 & 5   & 1.3  & 323  & 70 & 64.5 & 13   & 5.9  & 35.9 & 8.1  & 7.6  & -0.15 & 12.5 & V &  \\
212754 & 110785 & F7V   & 0.52  & 6.8 & 1.37 & 67   & 19 & 2083 & 508  & 5.8  & 56.7 & 16.4 & 7.3  & 0.16  & 14.9 & V & $\dagger$ \\
       &        &       &       &     &      &      &    & 39.4 & 11.8 &      &      &      &      &       &      & V &  \\
214454 & 111674 & A9V   & 0.24  & 96  & 2.02 & 1331 & 15 & 395  & 133  & 42.2 & 2256 & 717  & 105  & -0.19 & 150  & V &  \\
215588 & 112324 & F5V   & 0.388 & 13  & 1.26 & 1829 & 57 & 82.6 & 19.5 & 7.8  & 132  & 27.7 & 14.9 & -0.27 & 22.1 & V &  \\
215648 & 112447 & F6V   & 0.49  & 9   & 1.34 & 769  & 18 & 45.4 & 12.7 & 6.1  & 30.4 & 8.4  & 8.7  & -0.16 & 14   & V &  \\
216385 & 112935 & F6V   & 0.48  & 4.7 & 1.48 & 760  & 74 & 76.9 & 13.9 & 5.5  & 40.3 & 7.7  & 5.8  & -0.34 & 10.9 & V &  \\
218261 & 114096 & F6V   & 0.49  & 6   & 1.16 & 509  & 65 & 82   & 15.3 & 5.9  & 79.2 & 12.1 & 7.5  & -0.29 & 13.4 & V &  \\
218396 & 114189 & F0V   & 0.257 & 49. & 1.42 & 808  & 111& 4508 & 1072 & 36.4 &      &      &      &       &      & V &  \\
218470 & 114210 & F5V   & 0.44  & 8.8 & 1.48 & 2213 & 27 & 93.1 & 26.2 & 6.9  & 221  & 51.5 & 11.6 & 0.23  & 18.8 & V &  \\
220974 & 115770 & A6IV  & 0.16  & 98  & 1.99 & 1592 & 27 & 200  & 61   & 52.1 &28092 & 4746 & 130  & 0.22  & 153  & C &  \\
222368 & 116771 & F7V   & 0.5   & 7   & 1.38 & 3158 & 38 & 83.2 & 19.1 & 5.4  & 25.9 & 6.3  & 4.9  & 0.3   & 12.3 & V &  \\
222603 & 116928 & A7V   & 0.21  & 60  & 1.88 & 788  & 22 & 1556 & 380  & 26.3 & 4069 & 1336 & 64.5 & -0.49 & 85.5 & V &  \\
223731 & 117681 & F5V   & 0.44  & 31  & 1.33 & 778  & 21 & 188  & 48.5 & 17.7 & 500  & 151  & 42.5 & -0.21 & 46.9 & V &  \\
\hline
\end{longtable}
\twocolumn
\end{center}

\section{Sample selection}\label{app:select}

We selected the targets of our sample to try to keep it as representative as possible of AF MS dwarfs more massive than the Sun, while making it appropriate for accurate RV measurements with \safir. In terms of spectral type, we selected stars between A0 and $\sim$F8 types. The A0 limit corresponds roughly to the earliest spectral type for which close (sma $\lesssim$ 2.5 au) GP and/or BD companions are detectable with \safir~RV, while the F7-F9 limit roughly corresponds to the earliest spectral types for which the traditional cross-correlation technique with a binary mask gives accurate RV measurements. In terms of distance to the Sun ($d$, taken from the Hipparcos catalog), we set two cutoffs, both to obtain a statistically significant sample of nearby stars and to keep roughly the same proportion of A- and F-type targets: we set $d \leq$ 56 and $d \leq$ 33 parsecs for A- and F-type stars, respectively. To keep only objects of dwarf nature, we selected stars with an absolute magnitude within 2.5 mag from
the Main Sequence \citep{lagrange09}.\\

Then, we removed confirmed spectroscopic binaries of SB2 nature, close ($\leq$ 5 arcseconds) visual binaries and targets with highly variable luminosity, to keep targets observable in RV. Finally, we also removed confirmed $\delta$~Scuti and $\gamma$~Doradus pulsators, and chemically peculiar Am/Ap stars that are often associated with spectroscopic binaries \citep{lagrange09}. This selection leads to a visible dichotomy between our A- and F-type targets (as seen in Fig.~\ref{sample}), due to the removal of a significant number of late A and early F stars belonging to the instability strip. We initially ended up with $\sim$320 stars. We then restrained our effective \sophie~observing sample to 125 of these more than 300 targets (without any selection prior), to be able to fully cover the 1 to 1~000-day periodicity space for all targets during the \sophie~observations. These 125 targets can be divided into 72 stars with a mass $\leq$ 1.5 \Msun~and 53 stars with a mass $>$ 1.5 \Msun.

\section{Occurrence rate computation}\label{app:occ_rate}

For a given sample with $N$ stars, the search completeness function $C$($P$, \msini) gives the fraction of targets for which we can rule out a planet with a period $P$ and a minimal mass \msini~given the computed detection limits \citep[see \eg][]{howard11}. At a given ($P$, \msini) node, the completeness $C$ is defined as:
\begin{equation}
C = \frac{1}{N} \ \sum_{i = 1}^{N} \delta_{i}
\end{equation}
Here, for each target ($i$) of our sample without a detected planet, we have $\delta_{i} = 1$ if the computed detection limit at this ($P$, \msini) node is $<$ \msini~and $\delta_{i} = 0$ otherwise. We compute $C$($P$, \msini) for all the targets of our \sophie~+\harps~combined sample without a detected planet and with more than 10 spectra (219 targets overall). We compute the search completeness function using the detection limits computed from RV corrected from binary trends or RV-BIS correlations where appropriate.\\

Then, let’s consider a ($P$, \msini) domain ranging from $P_{\rm 1}$ to
$P_{\rm 2}$ in periods and from $m_{\rm p1} \sin{i}$ to $m_{\rm p2} \sin{i}$ in minimal masses, the ($P$, \msini) grid being the same as used for the detection limit computation. The number of systems with detected planets within this domain is $n_{\rm det}$ (\ie~a system with two planets within the same given domain counts for one). To derive the search completeness of our sample over $D$, we have to sum the search completeness function $C$($P$, \msini) over the period and the minimal mass ranges:
\begin{equation}
C_{D} = \frac{\displaystyle\sum_{P_{\rm 1}}^{P_{\rm 2}} \ \displaystyle\sum_{m_{\rm p1} \sin{i}}^{m_{\rm p2} \sin{i}} (\frac{1}{N} \displaystyle\sum_{i = 1}^{N} \delta_{i}) \ \mathrm{d}P \ \mathrm{d}(m_{\rm p} \sin{i})}{\displaystyle\sum_{P_{\rm 1}}^{P_{\rm 2}} \ \displaystyle\sum_{m_{\rm p1} \sin{i}}^{m_{\rm p2} \sin{i}} 1 \ \mathrm{d}P \ \mathrm{d}(m_{\rm p} \sin{i})}
\end{equation}
For a given domain $D$ with a search completeness $C_{D}$ and $n_{\rm det}$ systems with planets, the search completeness may be far from 100\%. It is thus necessary to estimate the number $n_{\rm miss}$ of potentially missed planets in order to correct for this incompleteness \citep{howard11}. We define $n_{\rm miss}$ as follows:
\begin{equation}
n_{\rm miss} = n_{\rm det} \times (\frac{1}{C_{D}} -1)
\end{equation}
Now we can derive the GP occurrence rate and its 1$\sigma$ and 2$\sigma$ uncertainties. We use binomial statistics, computing the binomial distribution of $n_{\rm det}$ GP systems among the $N$ targets. Given the probability $p$ of having one (or more) GP around one given star, we compute the probability $f$ of drawing drawing $n_{\rm det}$ systems among the $N$ stellar targets of our sample as follows:
\begin{equation}
f (n_{\rm det}, N, p) = \left(\! \begin{array}{c} N \\ n_{\rm det} \end{array} \!\right) \times \ p^{n_{\rm det}} \times \ (1 - p)^{N-n_{\rm det}}
\end{equation}
The result is the probability density function (PDF) of the GP occurrence rate (\ie, probability density versus $f$) in the given ($P$, \msini) domain. The GP occurrence rate is thus equal to the probability $f$ for which the PDF is the highest, multiplied by ($n_{\rm det} + n_{\rm miss}$)/$n_{\rm det}$ to account for the missed GP (see illustration in Paper IX). To derive the 1$\sigma$ or 2$\sigma$ uncertainties, we integrate the PDF and compute its standard deviations at $\sim$68 or $\sim$95\%, respectively.\\

In the case of a domain with no detected GP (\ie~$n_{\rm det} = 0$), we (re-)define $n_{\rm det}^{'} = 1$, $n_{\rm miss} = 1$/$C_{D}$ − 1, and compute the corresponding occurrence rate. The obtained value is then an upper limit on the real GP occurrence rate.

\section{Targets of peculiar interest: detailed plots}\label{app:details}

\begin{figure*}[ht!]
\centering
\includegraphics[width=1\hsize]{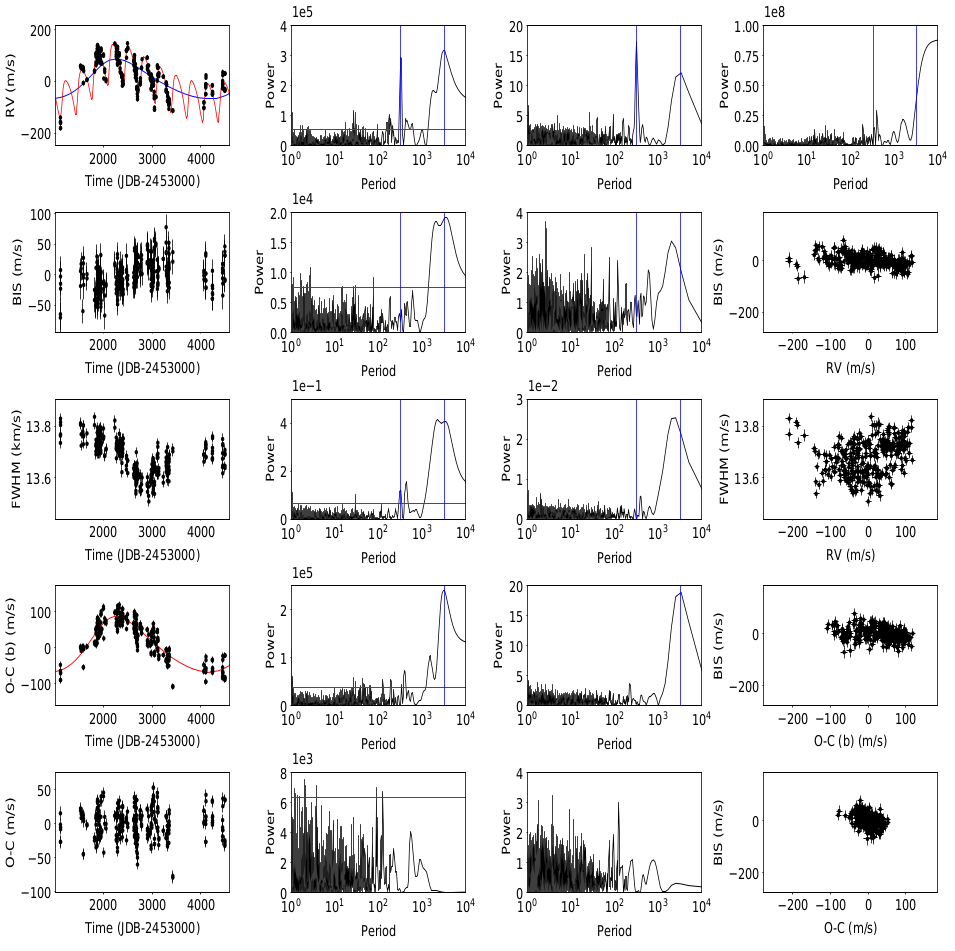}
\caption{HD~113337 spectroscopic data. {\it Top row, from left to right}: RV time series, Lomb-Scargle RV periodogram, CLEAN RV periodogram, and observational window function. The 2-planet Keplerian fit is overplotted ({\it red curve}) on the RV, as well as the Keplerian fit of the second candidate planet only ({\it blue curve}). {\it Second and third rows}: BIS and FWHM time series, corresponding Lomb-Scargle and CLEAN periodograms, and correlations with the RV data. {\it Fourth row}: RV residuals from planet b, corresponding Lomb-Scargle and CLEAN periodograms, and correlation with the BIS. The Keplerian fit of the candidate second planet is overplotted on the RV residuals ({\it red curve}). {\it Fifth row}: the same, for the RV residuals of the 2-planet Keplerian fit. {\it Note}: on all Lomb-Scargle periodograms, the $1\%$ false-alarm probability (FAP) is plotted in red. On all periodograms, the periods of planet b and/or candidate planet c are plotted in blue.}
\label{hd113337}
\end{figure*}

\begin{figure*}[ht!]
  \centering
\includegraphics[width=1\hsize]{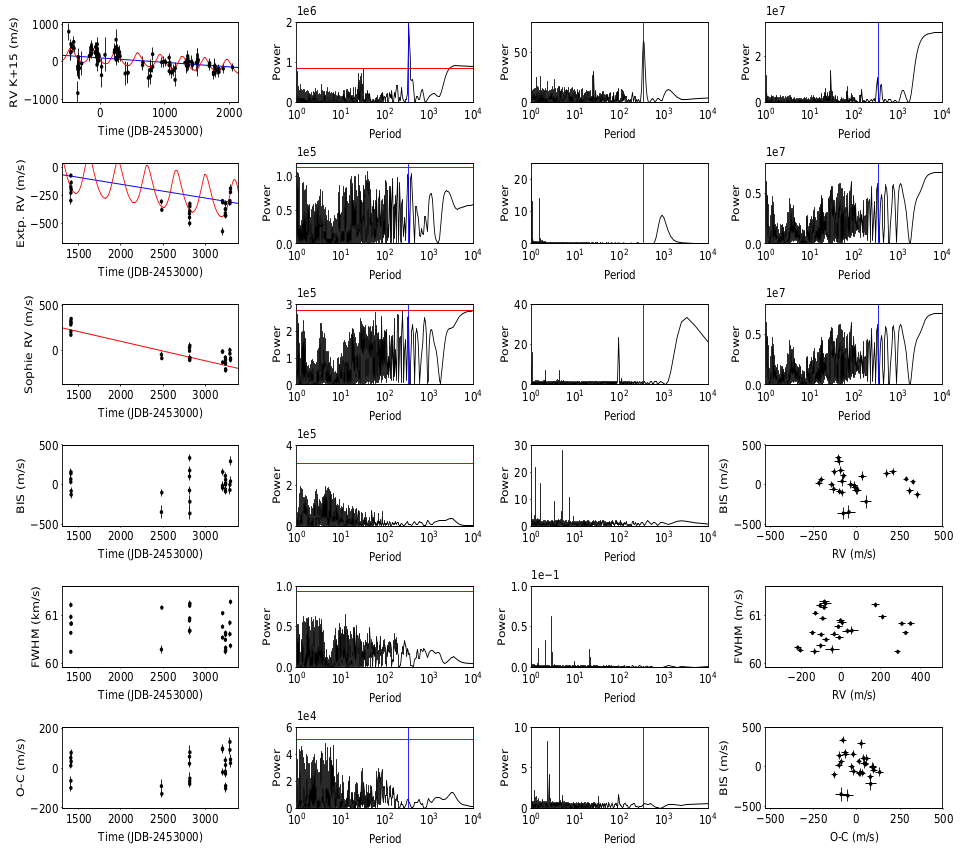}
\caption{HD~16232 spectroscopic data. {\it First row} ({\it from left to right}): HD~16232 RV time series from K15, corresponding Lomb-Scargle and CLEAN periodograms, and window function. The Keplerian fit of HD~16232b we obtained with {\it yorbit} based on the K15 data is overplotted ({\it red}) to the RV, as well as the linear drift ({\it blue}). {\it Second row}: the same for the RV expected from the Keplerian+linear fit of K15 data extrapolated at the epochs of our \sophie~observations, plus added noise. {\it Third row}: the same for the \sophie~RV time series. The linear fit we computed with {\it yorbit} based on the \sophie~RV is overplotted to the RV ({\it red solid line}). Note that each RV data set (\sophie~and K15 data) was independently centered on 0, hence the global RV offset between the two. {\it Fourth and fifth rows}: \sophie~BIS and FWHM time series, corresponding periodograms and correlations with \sophie~RV. {\it Sixth row}: Residuals from the linear fit to the \sophie~RV, corresponding periodograms and correlation to \sophie~BIS. Note: on the RV periodograms, HD~16232b period is overplotted in blue.}
\label{hd16232}
\end{figure*}

\begin{figure*}[t!]
  \centering
\includegraphics[width=1\hsize]{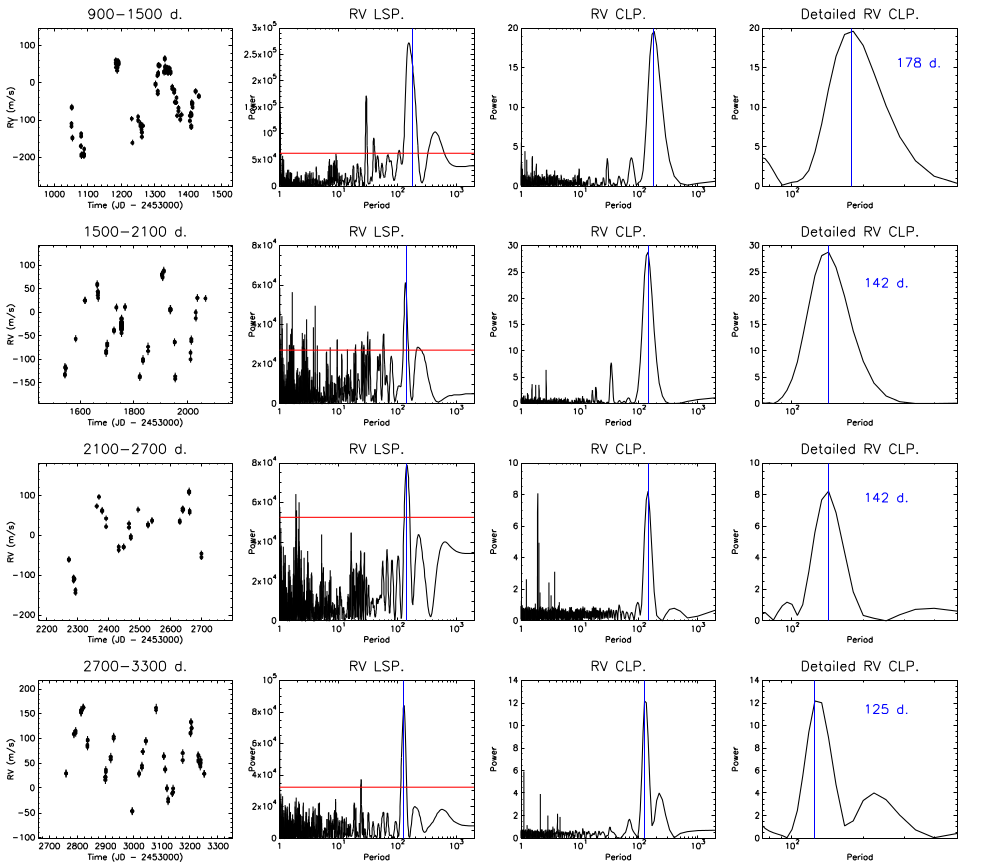}
\caption{Detail of $\theta$ Cyg \sophie~RV variations over four ({\it from top to bottom}) 600-day slices. {\it From left to right}: RV variations over the time slice, RV Lomb-Scargle periodogram (with 1\%~FAP in {\it red}), RV CLEAN periodogram, detail of the CLEAN periodogram. For each time slice, the main RV periodicity deduced from the CLEAN periodogram is overplotted in red.}
\label{theta_cyg_intv}
\end{figure*}

Here, we display in details the spectroscopic data of our targets with planetary companions (HD~113337, HD~16232) or that exhibit complex RV variations (HD~185395, HD~191195). These plots include specifically the Lomb-Scargle and CLEAN periodograms of the different observables. Regarding $\theta$~Cyg, we explored in more details the $\sim$150-day RV periodicity in order to try to understand why no Keplerian model can be correctly fitted. We separated the \sophie~RV data set over four 600-day slices (with more or less the same amount of data) and computed the corresponding Lomb-Scargle and CLEAN periodograms (Fig.~\ref{theta_cyg_intv}). We find quite astonishingly that the main mid-term RV periodicity is not stable at around 150 days, but decreases from $\sim$180 to $\sim$120 days. Furthermore, if looking at the \elodie~RV data set (that extends from BJD - 2453000 $\simeq - 100$ to $\simeq$ 1000, \ie~before the \sophie~observations), we find this time a periodicity of $\sim$130 days. A similar process applied to the \cite{howard16} RV data set (two $\sim$2000-day slices) give a main mid-term RV periodicity of 260 or 150 days, depending on the selected time slice. We thus conclude that the observed $\theta$~Cyg RV variations around $\sim$150 days are probably not really periodic but rather quasi-periodic. This does not favour the hypothesis of a substellar companion for which the RV periodicity would have to be more stable.

\begin{figure*}[t!]
\centering
\includegraphics[width=1\hsize]{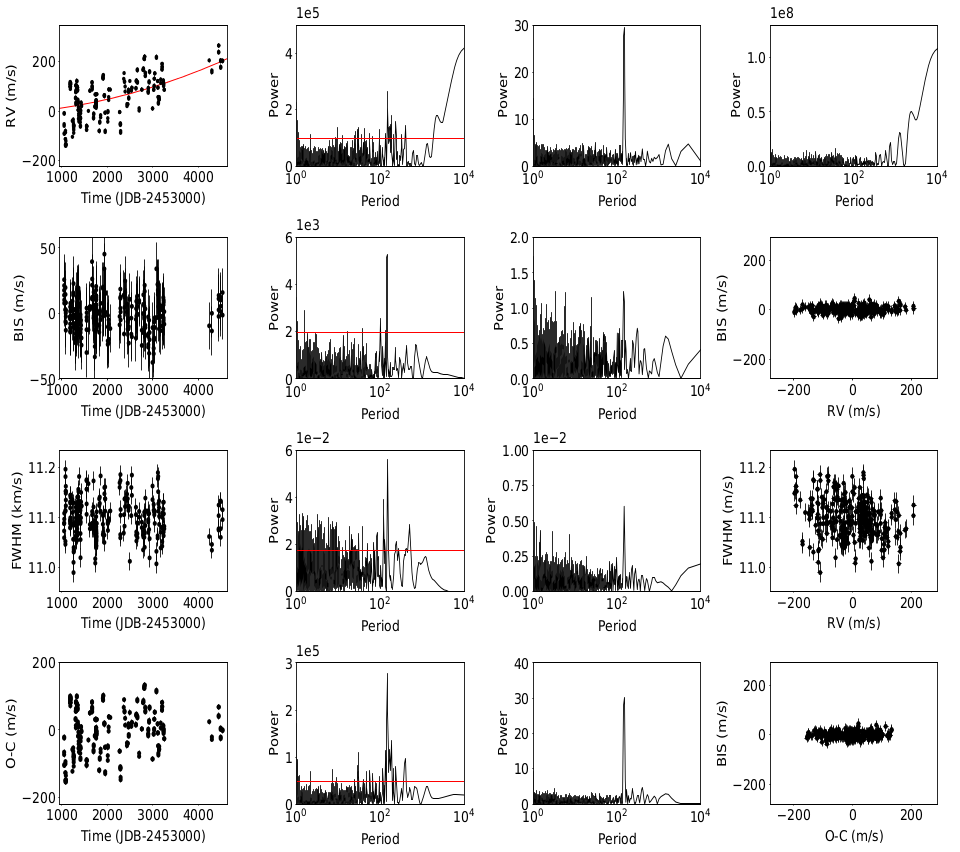}
\caption{$\theta$ Cyg spectroscopic data. {\it Top row, from left to right}: RV time series, Lomb-Scargle and CLEAN periodograms of the RV, and window function of the observations. The RV long-term quadratic trend is overplotted to the RV ({\it red solid curve}). {\it Second and third rows}: BIS and FWHM time series, corresponding Lomb-Scargle and CLEAN periodograms, and correlations with the RV. {\it Fourth row}: RV residuals from the quadratic fit, Lomb-Scargle and CLEAN periodograms of the RV residuals, and correlation of the RV residuals with the BIS.}
\label{hd185395}
\end{figure*}

\begin{figure*}[t!]
\centering
\includegraphics[width=1\hsize]{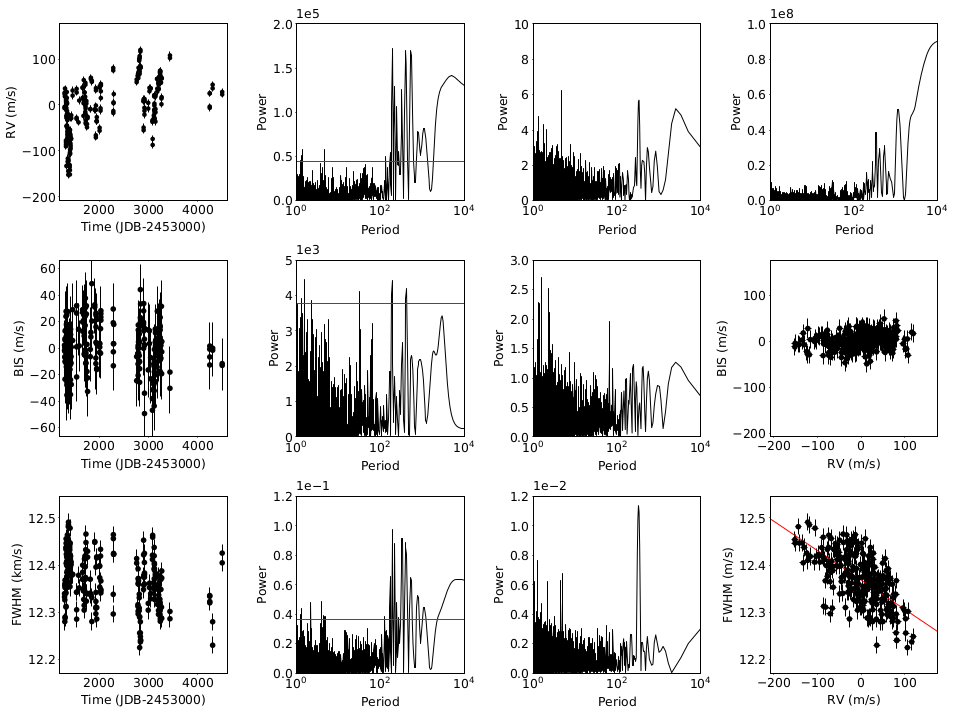}
\caption{HD~191195 spectroscopic data. {\it Top row, from left to right}: RV time series, Lomb-Scargle and CLEAN periodograms of the RV, and window function of the observations. {\it Second and third rows}: BIS and FWHM time series, corresponding Lomb-Scargle and CLEAN periodograms, and correlations with the RV. The RV-FWHM correlation is overplotted in red.}
\label{hd191195}
\end{figure*}

\end{appendix}

\end{document}